\documentclass[pra,reprint,a4paper]{revtex4-2}

\usepackage[T1]{fontenc}
\usepackage[utf8]{inputenc}
\newcommand{\latin}[1]{{\emph{#1}}}

\usepackage{dsfont} %Police pour écrire les ensembles usuels
\newcommand{\identite}{\ensuremath{\mathds{1}}} % Identité avec 1 avec double barre

\newcommand{\bigO}[1]{\ensuremath{\mathop{}\mathopen{}O\mathopen{}\left(#1\right)}} %dominé
\usepackage{graphicx}
\usepackage{color}
\usepackage{physics}
\usepackage{SIunits}
\usepackage{mathtools}

\usepackage[colorlinks]{hyperref}
\newcommand{\hypergeom}[1]{\operatorname{{}_2F_1}\left(#1\right)}
\newcommand{\aref}[1]{\hyperref[#1]{Appendix\,\ref{#1}}}
\def\figureautorefname~#1\null{%
  Fig.\,#1\null
}
\def\sectionautorefname~#1\null{%
  Sec.\,#1\null
}
\def\equationautorefname~#1\null{%
	Eq.\,(#1)\null
}

\hypersetup{%
pdftitle={Scheme for the generation of hybrid entanglement between time-bin and wavelike encoding},
pdfauthor={Gouzien, Élie and Brunel, Floriane and Tanzilli, Sébastien and D'Auria, Virginia},
pdfsubject={},
pdfkeywords={hybrid entanglement, quantum optics, quantum information}%
}

\begin{document}

\title{Scheme for the generation of hybrid entanglement between time-bin and wavelike encodings}

\author{Élie Gouzien}
\affiliation{Université Côte d'Azur, CNRS, Institut de Physique de Nice (INPHYNI), Parc Valrose, 06108 Nice Cedex 2, France}
\author{Floriane Brunel}
\affiliation{Université Côte d'Azur, CNRS, Institut de Physique de Nice (INPHYNI), Parc Valrose, 06108 Nice Cedex 2, France}
\author{Sébastien Tanzilli}
\affiliation{Université Côte d'Azur, CNRS, Institut de Physique de Nice (INPHYNI), Parc Valrose, 06108 Nice Cedex 2, France}
\author{Virginia D'Auria}
\email{virginia.dauria@inphyni.cnrs.fr}
\affiliation{Université Côte d'Azur, CNRS, Institut de Physique de Nice (INPHYNI), Parc Valrose, 06108 Nice Cedex 2, France}

\date{\today}

%\doi{10.1103/PhysRevA.102.012603}

\begin{abstract}
We propose a scheme for the generation of hybrid states entangling a single-photon time-bin qubit with a coherent-state qubit encoded on phases.
Compared to other reported solutions, time-bin encoding makes hybrid entanglement particularly well adapted to applications involving long-distance propagation in optical fibers.
This makes our proposal a promising resource for future out-of-the-laboratory quantum communication.
In this perspective, we analyze our scheme by taking into account realistic experimental resources and discuss the impact of their imperfections on the quality of the obtained hybrid state.
\end{abstract}

%\keywords{hybrid entanglement, quantum, optics}

\maketitle

\section{Introduction}

Over the past years, quantum optics information has traditionally followed two distinct approaches, naturally stemming from light wave-particle complementarity~\cite{FurusawaNP2015Hybriddiscretecontinuous}.
Discrete variable regime (DV) usually refers to weakly excited optical states, down to single photons, for which information is encoded on discrete spectrum observables such as the polarization or the number of photons~\cite{BlattNP2014Quantuminformationtransfer}.
Conversely, continuous variables regime (CV) relies on multiphoton optical states and to encodings on continuous spectrum observables such as amplitude and phase of a light field~\cite{LeeOSID2014ContinuousVariableQuantum}.
DV encoding is tolerant to losses and allows high-fidelity teleportation, while CV encoding permits deterministic state generation and unambiguous state discrimination~\cite{JeongPR2019Limitationsteleportingqubit}.

Recently, hybrid states entangling DV and CV encoding have been identified as a key tool to switch from one approach to the other and gather the benefits of both~\cite{FurusawaNP2015Hybriddiscretecontinuous, SoerensenPRL2010HybridLongDistance, JeongPRA2013deterministicquantumteleportation, SpillerJoO2017Hybridphotonicloss, JeongPRA2016Lossresilientphotonic, MilburnPRA2008Hybridquantumcomputation}.
This perspective has motivated an increasing number of theoretical works~\cite{JeongPRA2015Generationhybridentanglement, WangPRA2018Experimentallyfeasiblegeneration, LoockPRA2019Hybridquantumrepeater, Jeong2019Resourceefficientfault, JeongPRA2012Quantumteleportationparticlelike} as well as experiments on hybrid state generation~\cite{BelliniNP2014Generationhybridentanglement, LauratNP2014Remotecreationhybrid} or use in proof-of-principle quantum information protocols with single-rail or polarization encodings for the DV part~\cite{LvovskyPRL2017QuantumTeleportationDiscrete, LauratPRL2018DemonstrationEinsteinPodolsky, LauratO2018Remotepreparationcontinuous, LvovskyNC2018Entanglementteleportationpolarization}.
At the same time, practical quantum communication and networks will require the distribution of hybrid entanglement over long distances, where high losses or polarization instability can play a significant role.

This work addresses future applications of hybrid entanglement for long-distance operation over optical fiber links, by proposing a scheme for the generation of hybrid entangled states with time-bin encoding on their discrete variable part.
In the time-bin scheme, information is encoded on two generation or detection times for photons, usually labeled as ``early'' ($e$) and ``late'' ($l$) and generally obtained by exploiting a Franson interferometer~\cite{FransonPRL1989Bellinequalityposition}.
Compared to photon-number or polarization encodings, time bin allows one to comply in a better way with losses, is immune to polarization dispersion, and is particularly well adapted to quantum communication over long optical fibers~\cite{ZbindenRoMP2002Quantumcryptography}.
As a consequence, the realization of hybrid entanglement with time-bin encoding is extremely important in view of out-of-laboratory applications.
Our scheme permits one to generate the desired state in a heralded fashion without any postselection operation and relies, as enabling resources, on the interference between experimentally achievable states, \latin{i.e.\@} an optical Schrödinger cat state~\cite{GrangierN2007GenerationopticalSchroedinger, LvovskyNP2017EnlargementopticalSchroedingers}, and a time-bin entangled photon pair~\cite{TanzilliJoO2016Quantumphotonicstelecom}.
In order to comply with future practical realizations, it has been conceived so as to be experimentally feasible and fully compatible with existing fiber architectures and with realistic experimental resources, including nonideal heralding detectors~\cite{DAuriaPRA2018Quantumdescriptiontiming, HadfieldNP2009Singlephotondetectors} and input states~\cite{TanzilliJoO2016Quantumphotonicstelecom}.

In the following, we illustrate our scheme in detail.
For pedagogical reasons, in \autoref{seq:experience_proposee} we analyze it in the case of perfect input states and ideal heralding detectors.
In \autoref{seq:imparfait}, we investigate more realistic situations.
We start by considering for the heralding feature a minimum number of on-off single-photon detectors~\cite{DAuriaPRA2018Quantumdescriptiontiming, HadfieldNP2009Singlephotondetectors}, with no photon-number-resolving ability and finite efficiency.
We then examine the impact of realistic input states such as photon pairs generated by a nonlinear process along with vacuum and multiple pairs~\cite{TanzilliJoO2016Quantumphotonicstelecom} on the DV part, and a squeezed vacuum as an approximation of a Schrödinger cat on the CV one~\cite{LvovskyNC2018Entanglementteleportationpolarization}.
We show that our scheme is resistant to these experimental limitations by discussing the heralded state fidelity with respect to the targeted one.

\section{Proposed experiment\label{seq:experience_proposee}}
\subsection{Generation scheme of hybrid entanglement with time-bin DV encoding}
\begin{figure}[h]
  \centering
  \includegraphics[width=\linewidth]{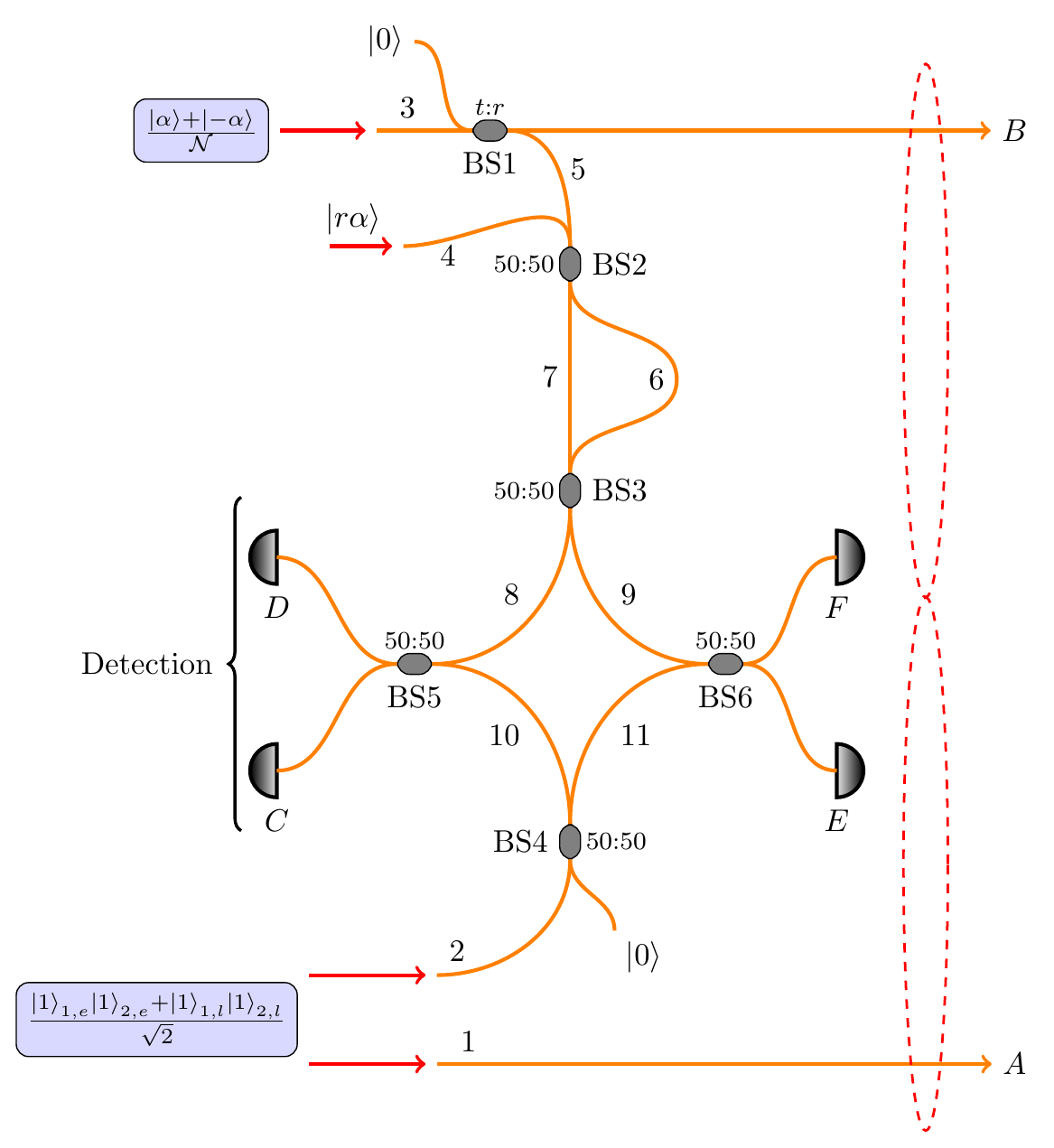}
  \caption{\label{fig:schema}Scheme for hybrid entanglement generation with time-bin encoding on the discrete variable part.}
\end{figure}

The aim of our work is to generate a DV-CV hybrid state, entangling a DV time-bin qubit with a bright CV qubit encoded on two coherent states with opposite phases.
We define such a target state as
\begin{equation}\label{eq:output_state}
\ket{\varphi} = \frac{\ket{1}_{A,e}\ket{+\alpha_f}_B - \ket{1}_{A,l}\ket{-\alpha_f}_B}{\sqrt{2}},
\end{equation}
where, for the DV part (mode~$A$), $e$ and $l$ stand for the early and late time bins and the $\ket{1}_{A,i}$ indicates a single photon state in the temporal mode~$i$.
For the CV part (mode~$B$), $\ket{\pm\alpha_f}_B$ is a coherent state of amplitude $\pm \alpha_f$.
To simplify the notation, in \autoref{eq:output_state} we omitted the vacuum states $\ket{0}_{A,l}$ and $\ket{0}_{A,e}$ that multiply the CV terms $\ket{+\alpha_f}_B$ and $\ket{-\alpha_f}_B$, respectively.
The density matrix associated with this state is $\ketbra{\varphi}$.

The experimental scheme at the heart of our proposal is presented in \autoref{fig:schema}.
Conceptually, it requires pure CV and DV states as initial resources and an interferometric scheme able to entangle them upon the result of a suitable measurement operation.
In this section we discuss the case of ideal inputs, whereas more realistic states from the experimental point of view will be treated in the following.
Note that the entire realization refers to the case of an experiment operated in pulsed regime, as required for the time-bin encoding.

At the CV input, labeled as mode~$3$ in the figure, we consider an even Schrödinger cat.
In terms of displacement operator $\hat{D}_{3}(\pm\alpha)$, this state can be written as
\begin{equation}\label{cat}
\ket{\text{cat+}}_{3} = \frac{\hat{D}_{3}(\alpha) + \hat{D}_{3}(-\alpha)}{\mathcal{N}} \ket{0},
\end{equation}
where we recall that $\hat{D}_{3}(\pm\alpha)\ket{0}=\ket{\pm\alpha}_{3}$, \latin{i.e.\@} a coherent state of amplitude $\pm\alpha$ and $\mathcal{N} = \sqrt{2} \sqrt{1 + e^{-2 {\abs{\alpha}}^2}}$~\cite{GrangierN2007GenerationopticalSchroedinger}.

Following an approach analogous to previously reported experiments~\cite{LauratNP2014Remotecreationhybrid}, the CV input state $\ket{\text{cat+}}_{3}$ is sent to an unbalanced beam splitter (BS1) with field reflection and transmission coefficients $r$ and $t$, respectively.
We note that unbalanced and variable beam splitters are easily available in both bulk and fiber configurations~\cite{LauratNP2014Remotecreationhybrid, LvovskyNC2018Entanglementteleportationpolarization, TanzilliJoO2016Quantumphotonicstelecom}.
After the BS1, the state reads
 \begin{equation}\label{eq:after_BS1}
\ket{\psi}_{B,5}=\frac{\hat{D}_{B}(t \alpha) \hat{D}_{5}(r \alpha)
      +  \hat{D}_{B}(-t \alpha) \hat{D}_{5}(-r \alpha)}{\mathcal{N}}
\ket{0}.%_{5, B}
\end{equation}
Mode~$B$ represents the CV part of our final hybrid state.
Mode~$5$ is directed towards the interferometric part of the scheme so as to be subsequently mixed with the DV input state.
This action actually permits bridging the CV and the DV parts of the state.

More precisely, light in mode~$5$ is sent to an unbalanced Mach-Zehnder interferometer, where it is mixed with a coherent state of amplitude $r\alpha$ at the input of a 50:50 beam splitter (BS2).
As for pure DV experiments~\cite{TanzilliJoO2016Quantumphotonicstelecom}, the length of the interferometer arms, $6$ and $7$, define the late and early time bins required for the encoding.
The state at the output of BS2 can be easily computed by recalling that given two input coherent states $\ket{\gamma}_{5}$ and $\ket{\gamma'}_{4}$, the output of a balanced beam splitter can be obtained by using the relation $\hat{D}_{5}(\gamma) \hat{D}_{4}(\gamma') = \hat{D}_{6}(\frac{\gamma - \gamma'}{\sqrt{2}}) \hat{D}_{7}(\frac{\gamma + \gamma'}{\sqrt{2}})$~\cite{ParisPLA1996Displacementoperatorbeam}.
For the case under examination, $\gamma = \pm r \alpha$ and $\gamma' = r \alpha$.
Accordingly, due to the interference with the coherent state on mode~$4$, depending on the sign of $\pm r \alpha$ on mode~$5$, light is routed only towards mode~$6$ (for $-r \alpha$) or mode~$7$ (for $r \alpha$).
As a consequence, right before the balanced beam splitter BS3, we obtain the state
%\begin{equation}
\begin{multline}\label{eq:after_BS2}
\ket{\psi}_{B,6,7}=\frac{1}{\mathcal{N}}(\hat{D}_{B}(t \alpha)\hat{D}_{6, l}(\sqrt{2}r \alpha) +\\
\hat{D}_{B}(-t \alpha)\hat{D}_{7, e}(-\sqrt{2} r \alpha))
\ket{0}.%_{6, 7, B}
\end{multline}
%\end{equation}
In the latter expression we use a double index notation to explicitly recall the temporal mode associated with spatial modes $6$ and $7$.

To obtain the desired hybrid state, light coming from the CV part, and prepared in time-bin modes by the Mach-Zehnder interferometer, must be mixed with the discrete variable part.
This is provided by a pair of time-bin entangled photons launched in the input spatial modes $1$ and $2$~\cite{TanzilliJoO2016Quantumphotonicstelecom}:
\begin{equation}\label{eq:input_pair}
\ket{\xi}_{1,2}=\frac{\ket{1}_{1,e} \ket{1}_{2,e} + \ket{1}_{1,l} \ket{1}_{2,l}}{\sqrt{2}}.
\end{equation}

One of the photons is directly routed towards output mode~$A$ and represents the DV part of the hybrid state.
Its twin, on mode~$2$, is sent to the balanced beam splitter BS4 so as to be spatially mixed with the two outputs of the Mach-Zehnder interferometer thanks to BS5 and BS6 (see \autoref{fig:schema}).
Balanced beam splitters BS5 and BS6 erase the ``which path'' information, right before the heralding detectors (in spatial modes $C$, $D$, $E$, $F$), such that a given click event from one of the four detectors cannot be attributed to light coming from a certain origin (\latin{i.e.\@} from the CV or the DV part).
Right before the detectors, the state reads
\begin{widetext}
\begin{multline}\label{eq:etat_detecteurs}
\ket{\psi^{(1)}} =
    \frac{1}{2 \sqrt{2} \mathcal{N}} \Bigg[ \\
    \begin{aligned}[t]
      &\phantom{{}+{}} \ket{+ t \alpha}_B \ket{1}_{A,e}
         \left\lbrace
         \hat{D}_{C, l}\left(+\frac{r \alpha}{\sqrt{2}}\right)
         \hat{D}_{D, l}\left(-\frac{r \alpha}{\sqrt{2}}\right)
         \hat{D}_{E, l}\left(-\frac{r \alpha}{\sqrt{2}}\right)
         \hat{D}_{F, l}\left(+\frac{r \alpha}{\sqrt{2}}\right)\ket{0}_{l}
         \right\rbrace
         \left[\ket{1}_{C,e}+\ket{1}_{D,e}+\ket{1}_{E,e}+\ket{1}_{F,e}\right] \\
      &+ \ket{-t \alpha}_B \ket{1}_{A,e} \ket{0}_{l}
         \left\lbrace
         \hat{D}_{C, e}\left(-\frac{r \alpha}{\sqrt{2}}\right)
         \hat{D}_{D, e}\left(+\frac{r \alpha}{\sqrt{2}}\right)
         \hat{D}_{E, e}\left(-\frac{r \alpha}{\sqrt{2}}\right)
         \hat{D}_{F, e}\left(+\frac{r \alpha}{\sqrt{2}}\right)
         \left[\ket{1}_{C,e}+\ket{1}_{D,e}+\ket{1}_{E,e}+\ket{1}_{F,e}\right] \right\rbrace \\
      &+ \ket{+t \alpha}_B \ket{1}_{A,l} \ket{0}_{e}
          \left\lbrace
          \hat{D}_{C, l}\left(+\frac{r \alpha}{\sqrt{2}}\right)
          \hat{D}_{D, l}\left(-\frac{r \alpha}{\sqrt{2}}\right)
          \hat{D}_{E, l}\left(-\frac{r \alpha}{\sqrt{2}}\right)
          \hat{D}_{F, l}\left(+\frac{r \alpha}{\sqrt{2}}\right)
          \left[\ket{1}_{C,l}+\ket{1}_{D,l}+\ket{1}_{E,l}+\ket{1}_{F,l}\right] \right\rbrace \\
      &+ \ket{-t \alpha}_B \ket{1}_{A,l}
          \left\lbrace
          \hat{D}_{C, e}\left(-\frac{r \alpha}{\sqrt{2}}\right)
          \hat{D}_{D, e}\left(+\frac{r \alpha}{\sqrt{2}}\right)
          \hat{D}_{E, e}\left(-\frac{r \alpha}{\sqrt{2}}\right)
          \hat{D}_{F, e}\left(+\frac{r \alpha}{\sqrt{2}}\right)\ket{0}_{e}
          \right\rbrace
          \left[\ket{1}_{C,l}+\ket{1}_{D,l}+\ket{1}_{E,l}+\ket{1}_{F,l}\right] \Bigg],
    \end{aligned}
\end{multline}
where $\ket{0}_l$ indicates $\ket{0}_{C,l}\ket{0}_{D,l}\ket{0}_{E,l}\ket{0}_{F,l}$ and analogously for $\ket{0}_e$.
We note that $\frac{r \alpha}{\sqrt{2}}$ corresponds to the amplitude of the light beam coming from the continuous variable part and reaching the detectors.
In the notation $\ket{\psi^{(1)}}$, label $1$ indicates that we consider exactly one photon pair on the discrete variable input.
\end{widetext}

Light in the four spatial modes $C$, $D$, $E$, $F$ is measured by using single-photon detectors; for each of them two temporal modes ($l$ and $e$) must be considered, thus leading to eight possible heralding modes.
The combination of their detection signals heralds the hybrid state on $A$ and $B$.

We will consider here the ideal case of photon-number-resolving detectors with perfect detection efficiency~\cite{JeongPRA2015Generationhybridentanglement,
      WangPRA2018Experimentallyfeasiblegeneration} and herald the hybrid state by the simultaneous detection of signals corresponding to the measurement of one photon on detector $E$ in the late time bin, one photon on detector $F$ in the early time bin, and no photon in the six remaining heralding modes.
The associated positive operator valued measurement (POVM)~\cite{LauratPRL2011QuantumDecoherenceSingle} reads
\begin{equation}\label{POVMideal}
\hat{\mathbf{\Pi}}^{\text{id}} =
       \ketbra{1}_{E,l}  \ketbra{1}_{F,e}
       \bigotimes\limits_i \ketbra{0}_{i},
\end{equation}
where the label $i$ indicates the heralding modes $(C,e)$, $(C,l)$, $(D,e)$, $(D,l)$, $(E,e)$ and $(F,l)$.
In the previous equation, identity is implicit on unmeasured channels $A$ and $B$.
As it can be seen, only the first and the fourth terms of \autoref{eq:etat_detecteurs} simultaneously contain light in the $(E,l)$ and the $(F,e)$ modes, and can lead to a detection event as described by $\hat{\mathbf{\Pi}}^{\text{id}}$.
We note that the detected photons are provided one by the DV and the other by the CV part of the scheme.
On the contrary, the second and the third terms contain light only in one of the two time bins and do not contribute to the announced states.
Accordingly, as desired, the chosen heralding strategy exactly leads to the target state of \autoref{eq:output_state}:
\begin{equation}\label{rhoideal}
\hat{\rho}^{\text{id}}
	= \dfrac{\Tr_{CDEF}\left[\hat{\mathbf{\Pi}}^{\text{id}} \ketbra{\psi^{(1)}}\right]}{\Tr\left[\hat{\mathbf{\Pi}}^{\text{id}} \ketbra{\psi^{(1)}}\right]}
	= \ketbra{\varphi},
\end{equation}
where the amplitude of the CV qubit is $\alpha_{f} = t\alpha$.
The target state is heralded with a probability:
\begin{equation}\label{Pideal}
P^{\text{id}}
	= \Tr\left[\hat{\mathbf{\Pi}}^{\text{id}} \ketbra{\psi^{(1)}}\right]
	= \frac{1}{16}\frac{\abs{r \alpha}^2 e^{-2\abs{r \alpha}^2}}
                    {1 + e^{-2\abs{\alpha}^2}}.
\end{equation}
As it can be seen, equality~\eqref{rhoideal} holds true for any choice of the fraction of light, $r \alpha$, that is subtracted from the CV part and mixed with the DV one.
However, as shown in \autoref{fig:fidelite_et_taux_compare} (dashed lines), the heralding probability is strongly affected by $r \alpha$.
For a fixed size of the initial cat state, \latin{i.e.\@} for a given $\alpha$, $P^{\text{id}}$ increases with $r \alpha$ up to an optimal value and then decreases for higher $r \alpha$ values.
This behavior is easily justified by the heralding choice.
For $r \alpha \to 0$, the fraction of the continuous variable part directed towards detection modes is extremely weak and the probability of obtaining a detection signal from both $(E,l)$ or $(F,e)$ is low.
On the contrary, for high $r \alpha$, detection of light on modes $C$, $D$, $(E,e)$ and $(F,l)$ has an important probability to be triggered by the continuous variable component of $\ket{\psi^{(1)}}$ and, as for state preparation we impose the absence of photons on these detectors, the overall heralding probability decreases.
We also stress that for increasing $r \alpha$, the condition of detecting only one photon on heralding modes $(E,l)$ or $(F,e)$ is no longer respected as multiple photon contributions become non-negligible.

\subsection{Hybrid states with time-bin encoding in long-distance applications\label{subseq:Qcomm}}
Hybrid DV-CV entangled states are essential for any quantum networks connecting disparate quantum devices based on CV or DV encodings~\cite{FurusawaNP2015Hybriddiscretecontinuous} and are at the heart of a new generation of quantum communication protocols~\cite{LoockPRA2019Hybridquantumrepeater, Jeong2019Resourceefficientfault, JeongPRA2012Quantumteleportationparticlelike}.
At the same time, these applications require the distribution to remote nodes of one or both parts of the hybrid states; it is thus crucial to discuss their robustness in the context of long-distance operation.
We will consider here the experimentally relevant case of fiber connections submitted to propagation losses and dispersion.

Loss effect can be modeled by inserting unbalanced BSs with reflection coefficients $r_{\text{CV}}$ and $r_{\text{DV}}$ on the paths of the CV or DV parts of the states (see \aref{Loss}).
In practical scenarios, the reflection coefficient goes with the propagation distance, $z$, as $\sqrt{1-e^{-\beta\text{z}}}$ where, in optical fibers, $\beta \approx \unit{0.2}{\deci\bel\per\kilo\meter}$ at its best~\cite{TanzilliJoO2016Quantumphotonicstelecom}.
Losses on the CV part of a hybrid entangled state lead to decoherence and result in a degradation of the state purity with off-diagonal terms (coherences) of the state density matrix exponentially scaling as $e^{-2\abs{r_{\text{CV}}\alpha_f}^2}$.
In addition, the state coherent amplitude is reduced as $\pm \sqrt{1-r_{\text{CV}}^2} \alpha_f$.
These limitations are independent of the chosen DV encoding and they are common to any hybrid CV-DV state submitted to losses in the CV part.
Conversely, striking advantages of time-bin encoding appear when considering loss and propagation effects on the DV part of the state.

Losses on the DV part transform $\ketbra{\varphi}$ into
\begin{multline}\label{rholoss}
\tilde{\rho}
	= t_{\text{DV}}^2 \ketbra{\varphi} \\
		+ r_{\text{DV}}^2 \ketbra{0}_{A}
			\frac{\ketbra{+\alpha_{f}}_{B} + \ketbra{-\alpha_{f}}_{B}}{2},
\end{multline}
where $t_{\text{DV}}^2= 1-r_{\text{DV}}^2$ and we kept the label~$A$ for the DV part after the loss-beam splitter and $B$ for the CV part.
An expression similar to $\tilde{\rho}$ can be derived starting from $\hat{\rho}^{\text{id}}$.
The first term of \autoref{rholoss} contains the initial state multiplied by $t_{\text{DV}}^2$ and unaffected by decoherence effects.
The second term, proportional to $r_{\text{DV}}^2$, is a nonentangled hybrid state with the DV part, $A$, in a vacuum state; accordingly, in practical applications, this term can be traced out by DV detection operation as it happens in pure DV regime.
As widely discussed in the literature~\cite{ZbindenRoMP2002Quantumcryptography}, dispersion effects on time-bin encoding play no significant role and can be easily compensated by standard modules.
No decoherence effect is thus observed on the state $\ketbra{\varphi}$ when its DV part is submitted to losses or propagates over long distances.
Remarkably, these relevant features are not available to other kinds of reported hybrid entanglement.
When their DV part propagates over long optical fibers, hybrid entangled states with single-rail (presence or absence of a single photon) or polarization DV encodings are strongly affected by losses and polarization dispersion.
As a consequence, they suffer from a decoherence effect going exponentially with the propagation distance (see \aref{Loss}).

\begin{figure}[h]
	\centering
	\includegraphics[width=\linewidth]{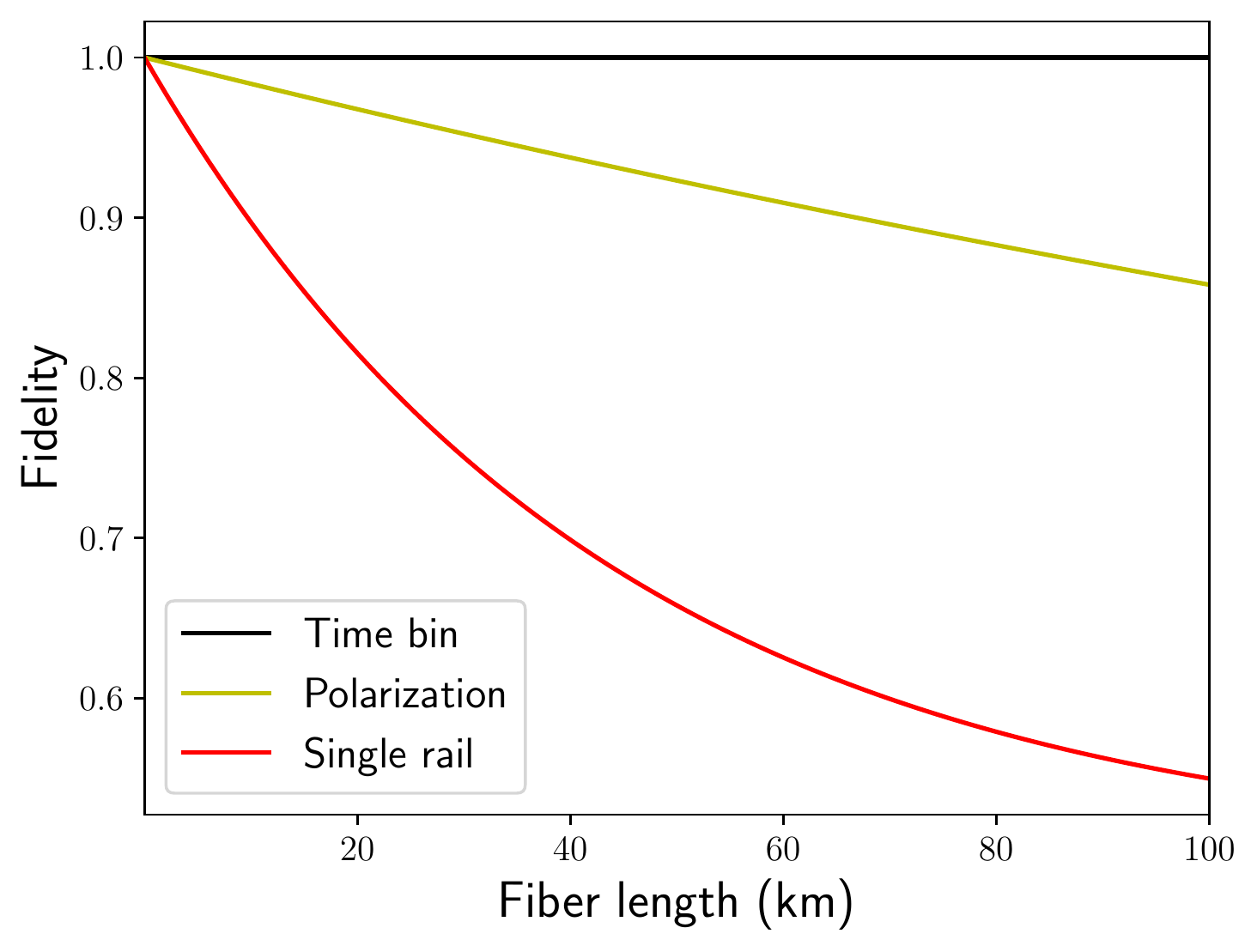}
	\caption{\label{Confronto}
		Fidelity with the target state of a CV qubit remotely prepared as a function of the distance traveled by the DV part of the hybrid state.
		The target state is an odd Schrödinger cat state.
		The cases of hybrid entangled states with time-bin, single-rail, and polarization DV encodings are compared.
		For the polarization dispersion, we considered the value reported in a recent out-of-the-laboratory experiment~\cite{UrsinPotNAoS2019Entanglementdistributionover}.
		We assumed a DV mode projected onto $\frac{\ket{1}_{A,l} + \ket{1}_{A,e}}{\sqrt{2}}$ for hybrid entanglement with time-bin DV encoding and similar projections for single-rail and polarization DV encodings (see \aref{Loss}).
		For the initial amplitude of the CV part of the hybrid states, we set $\alpha_f = 2$.
		The distance traveled by the CV part is taken as negligible for the three states.}
\end{figure}

The robustness of hybrid entangled states with time-bin encoding makes them excellent resources for protocols where local manipulation and detection of a part of the hybrid state are used to herald at distance a certain quantum operation.
As an example, we consider the simple case of remote preparation of a CV quantum state upon the results of the measurement of the DV part.
This scheme allows transferring quantum information between distant nodes, in a configuration where, in contrast with quantum teleportation, the sender has complete knowledge of the state to be communicated~\cite{LauratO2018Remotepreparationcontinuous}.
\hyperref[Confronto]{Figure\,\ref{Confronto}} reports the fidelity of the prepared CV state with the target one versus the distance traveled by the DV part.
Hybrid states with time-bin, single-rail and polarization DV encoding are compared.
Beside losses, pertinent dispersion effects have been considered.
As it can be seen, the fidelity obtained in the case of time bin is unaffected by the traveled distance, while the performances of hybrid state with single rail and polarization encoding degrade with it.
We stress that considerations similar to the ones that we discuss for this protocol can be easily extended to other heralded protocols such as teleportation or entanglement swapping, thus confirming the advantage of hybrid entangled with time-bin encoding for practical quantum communication.

\section{Robustness of the generation scheme against experimental limitations}\label{seq:imparfait}
\subsection{Imperfect detection and simplified heralding strategy\label{subseq:annonce_on}}

The ideal case discussed in the previous section relies on the possibility of detecting, \latin{i.e.\@}, two temporal modes for each of the four spatial modes $C$, $D$, $E$, $F$.
This can be seen from the shape of POVM~\eqref{POVMideal} that explicitly contains projectors over all the eight heralding modes.
In experiments, the strategy described in \autoref{seq:experience_proposee} would imply that each of the heralding detectors should be able to measure both early and late time bins.
This implies a separation between time bins greater than the detector dead time or, in alternative, the use of extra detectors so as to map each time bin in an additional spatial mode~\cite{LauratPRL2011QuantumDecoherenceSingle}.
These solutions have dramatic consequences for the experiment maximum operation rate and required overhead, respectively.
At the same time, we note that photon-number-resolving detectors, although often introduced in the literature on hybrid state~\cite{WangPRA2018Experimentallyfeasiblegeneration}, are hardly available off the shelf and, so far, most of the demonstrations involving some counting ability rely on complex spatial multiplexing strategies based on detector arrays~\cite{WalmsleyNJoP2009Measuringmeasurementtheory, SiegelIToAS2019Characterizationphotonnumber}.

\begin{figure}[h]
	\centering
	\includegraphics[width=\linewidth]{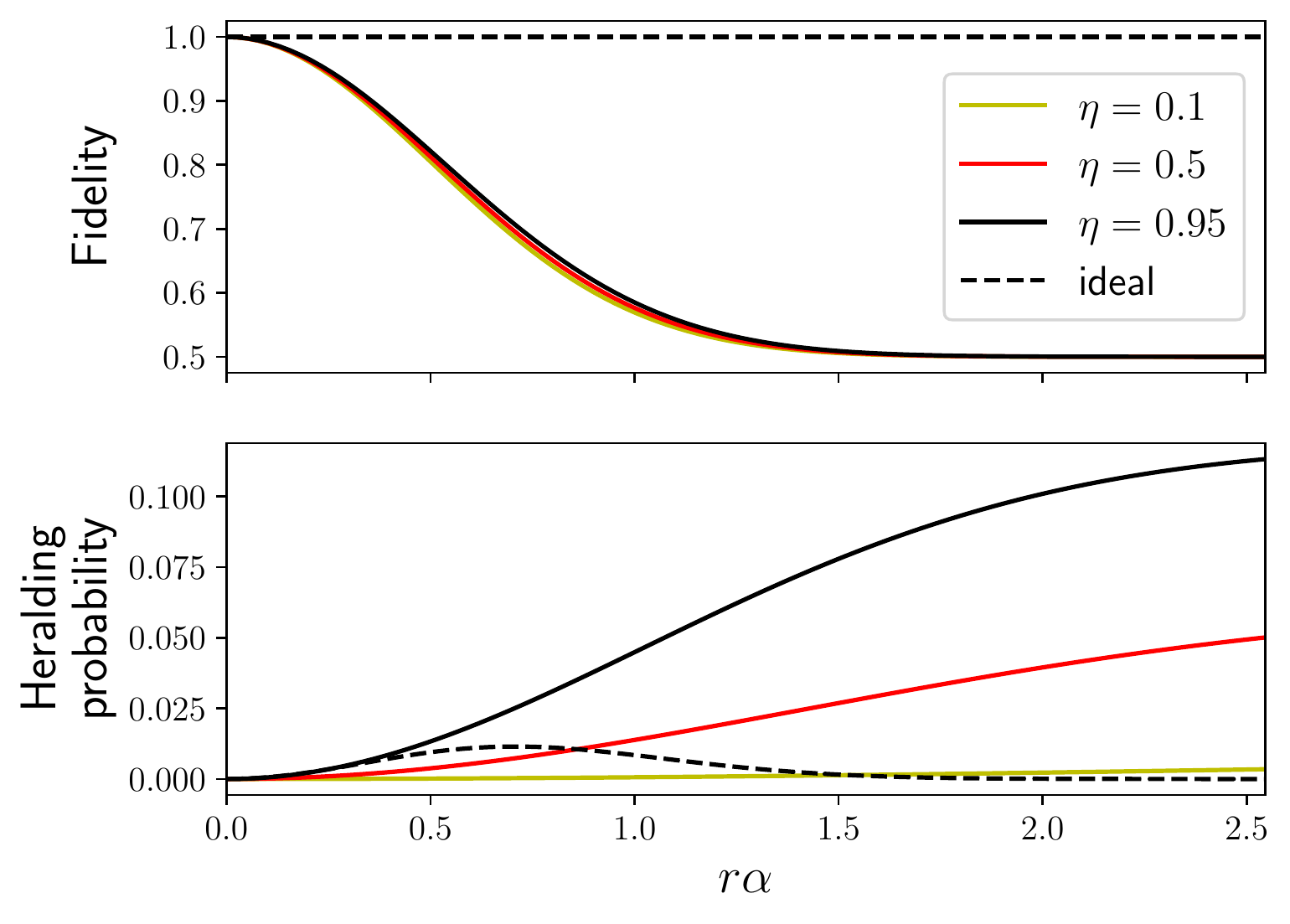}
	\caption{\label{fig:fidelite_et_taux_compare}
		Fidelity of the heralded hybrid state to the target state $\ketbra{\varphi}$ (top) and heralding probability (bottom) as a function of $r \alpha$, for the two proposed heralding protocols: $\hat{\mathbf{\Pi}}^{\text{id}}$ (dashed line) and $\hat{\mathbf{\Pi}}$ (plain lines).
		The ideal case is discussed in \autoref{seq:experience_proposee} and refers to perfect photon number resolving detectors with unitary efficiency.
		All other cases refer to on-off detector, whose quantum efficiencies, $\eta$, have been chosen according to typical experimental values~\cite{TanzilliJoO2016Quantumphotonicstelecom, ShieldsJoAP2015GigahertzgatedInGaAsInP, NamNP2013Detectingsingleinfrared}.
		For the input CV state, we set $\alpha = 2$.
		The fidelity and the product $r \alpha$ are adimensional quantities.}
\end{figure}

In view of future experimental realizations, in this section, we discuss our scheme in the case of a more realistic detection scenario.
A practical simplification of our protocol consists of heralding the target state only upon the detection signals from the late time bin in mode~$E$ and from the early time bin in mode~$F$, thus choosing to disregard all other six heralding modes that describe the state before the detector (see \autoref{eq:etat_detecteurs}).
By doing so, the heralded state will no longer correspond to $\hat{\rho}^{\text{id}}$.
At the same time, such an approach considerably reduces the number of detected modes from eight to two.
By doing so, only two gated detectors are required and, as each of them must detect only one temporal mode, dead times are no longer a limiting factor to the operation rate~\cite{DAuriaPRA2018Quantumdescriptiontiming}.
At the same time, we consider, at the heralding modes, on-off single-photon detectors, with nonunit detection efficiency $\eta$ and no photon-number-resolving ability~\cite{HadfieldNP2009Singlephotondetectors}.
These systems are able to provide only the two generic responses: on, \latin{i.e.\@} ``at least one photon has been detected'', and off, \latin{i.e.\@} ``no photon has been detected'', and they represent the vast majority of available single-photon counters.
Their action is described by the positive operators~\cite{DAuriaPRA2018Quantumdescriptiontiming, LauratPRL2011QuantumDecoherenceSingle,
WalmsleyNJoP2009Measuringmeasurementtheory}:
\begin{subequations}\label{POVMONOFF}
\begin{align}
\hat{\Pi}^{\textsc{off}}_{i}
  &= \sum\limits_{k=0}^{+\infty} {(1-\eta)}^k \ketbra{k}_{i}, \\
\hat{\Pi}^{\textsc{on}}_{i}
  &= \identite - \hat{\Pi}^{\textsc{off}},
\end{align}
\end{subequations}
where $i$ indicates the heralding mode under investigation and, only for the previous equations, $\ket{k}_{i}$ indicate a Fock state of mode~$i$ containing $k$ photons.

The measurement positive valued operator associated with the simplified heralding strategy and with on-off gated detectors reads
\begin{equation}
\hat{\mathbf{\Pi}} = \hat{\Pi}^{\textsc{on}}_{E, l} \otimes \hat{\Pi}^{\textsc{on}}_{F, e},
\end{equation}
identity being implicit on all six unmentioned modes (\latin{i.e.\@} $C$, $D$, $(E,e)$, $(F,l)$) as they are not measured.
We note, in particular, that, with this measurement strategy, the scheme no longer relies on projection on vacuum states.
Analogous to what is done in Eqs.\,\eqref{rhoideal}~and~\eqref{Pideal}, the heralded state can be computed from $\hat{\mathbf{\Pi}}$:
\begin{multline}
\hat{\rho}^{(1)}
   = \frac{1}{2}\Bigg\lbrace
           \ketbra{\alpha_f}{\alpha_f}_B
           \ketbra{1}{1}_{A,e} \\
         - \frac{\eta \frac{\abs{r \alpha}^2}{2} e^{- 2\abs{r \alpha}^2}}
                 {1 - e^{- \eta\frac{\abs{r \alpha}^2}{2}}}
            \left[\begin{multlined}
                    \ketbra{\alpha_f}{-\alpha_f}_B
                      \ketbra{1}{0}_{A,e} \ketbra{0}{1}_{A,l} \\
                    + \ketbra{-\alpha_f}{+\alpha_f}_B
                    \ketbra{0}{1}_{A,l} \ketbra{1}{0}_{A,e}
                  \end{multlined}\right] \\
         + \ketbra{-\alpha_f}{-\alpha_f}_B
            \ketbra{1}{1}_{A,l} \Bigg\rbrace.
\label{subeq:etat_annonce_on}
\end{multline}
Its associated heralding probability reads
\begin{align}
P^{(1)} = \frac{\eta}{8} \frac{1 - e^{- \eta\frac{\abs{r \alpha}^2}{2}}}{1 + e^{-2\abs{\alpha}^2}}, \label{subeq:taux_annonce_on}
\end{align}
where, as before, we used the notation $t\alpha=\alpha_f$.

The state given by~\eqref{subeq:etat_annonce_on} belongs to the qubit subspace generated by $\ket{+ \alpha_{f}}_B\ket{1}_{A,e}$ and $\ket{- \alpha_{f}}_B\ket{1}_{A,l}$, as the density matrix $\ketbra{\varphi}$, and correctly tends to it when $r \alpha \to 0$.
% Rappel : toutes les normes sont équivalentes car dimension finie.
This can be seen from the fidelity of the heralded state to the target state:
\begin{equation} \label{subeq:fidelite_annonce_on}
\mathcal{F}^{(1)}
	= \expval{\hat{\rho}^{(1)}}{\varphi}
	= \frac{1}{2} \left[1 + \frac{\eta \frac{\abs{r \alpha}^2}{2} e^{- 2\abs{r \alpha}^2}}
 	{1 - e^{- \eta\frac{\abs{r \alpha}^2}{2}}}\right].
\end{equation}

The heralding probability, $P_1$, and fidelity $\mathcal{F}^{(1)}$ are plotted in \autoref{fig:fidelite_et_taux_compare} as functions of $r \alpha$ for different detector efficiencies $\eta$.
In this regard we underline that the fidelity is a commonly adopted and pertinent criterium to check the validity of any scheme aiming at the generation of a given target state~\cite{JeongPRA2015Generationhybridentanglement, WangPRA2018Experimentallyfeasiblegeneration}.
% Dépendance en $\eta$
The efficiency $\eta$ has a clear impact on the heralding probability, $P_1$, on the order of $\eta^2$ in the limit of $r \alpha \to 0$.
Compared to the ideal case described in the previous section (dashed lines), the heralding probability is no longer decreasing for high $r \alpha$, as the heralding strategy is no longer sensitive to spurious events firing one of the disregarded modes $C$, $D$ and $(E,e), (F,l)$.
Nevertheless, the bad effect of multiphoton contributions coming from the CV part can be clearly seen on the state fidelity, $\mathcal{F}^{(1)}$, that decreases for increasing $r \alpha$.
Detection efficiency $\eta$ has little influence on the fidelity.
This weak dependency arises from the shape of the on operator in \autoref{POVMONOFF}.

In experiments, for a fixed input state $\ket{\text{cat+}}_3$, \latin{i.e.\@} for a given $\alpha$, and for given detector quantum efficiency $\eta$, the reflection coefficient $r$ can be chosen so as to reach a desired value for the fidelity.
As an example, we consider $\alpha = 2$ as in Refs.\,\cite{BelliniNP2014Generationhybridentanglement,
GrangierN2007GenerationopticalSchroedinger,
LauratNP2014Remotecreationhybrid,
LvovskyNP2017EnlargementopticalSchroedingers} and realistic single-photon detectors based on superconducting nanowire technology exhibiting $\eta = 0.95$~\cite{NamNP2013Detectingsingleinfrared} working in a gated operation mode.
With these parameters, a high target fidelity $\mathcal{F}^{(1)} = 0.99$ imposes $\frac{r \alpha}{\sqrt{2}} = 0.075$, thus giving $r \approx 0.053$.
Correspondingly, the heralding probability $P^{(1)}\approx6.4\cdot10^{-4}$.
Working with a \unit{1}{\giga\hertz} repetition rate laser~\cite{TanzilliLPR2015Ultrafastheralded}, and fast single-photon detectors compatible with such a fast operation regime, as those reported in Ref.\,\cite{NamNP2013Detectingsingleinfrared}, leads to a heralding rate of \unit{0.64}{\mega\hertz}.
Note that, as the fraction of light subtracted from the CV input is very small, the size of the CV part of the final hybrid state is $\alpha_f\approx2$.

\begin{figure}[!h]
	\centering
	\includegraphics[width=1\linewidth]{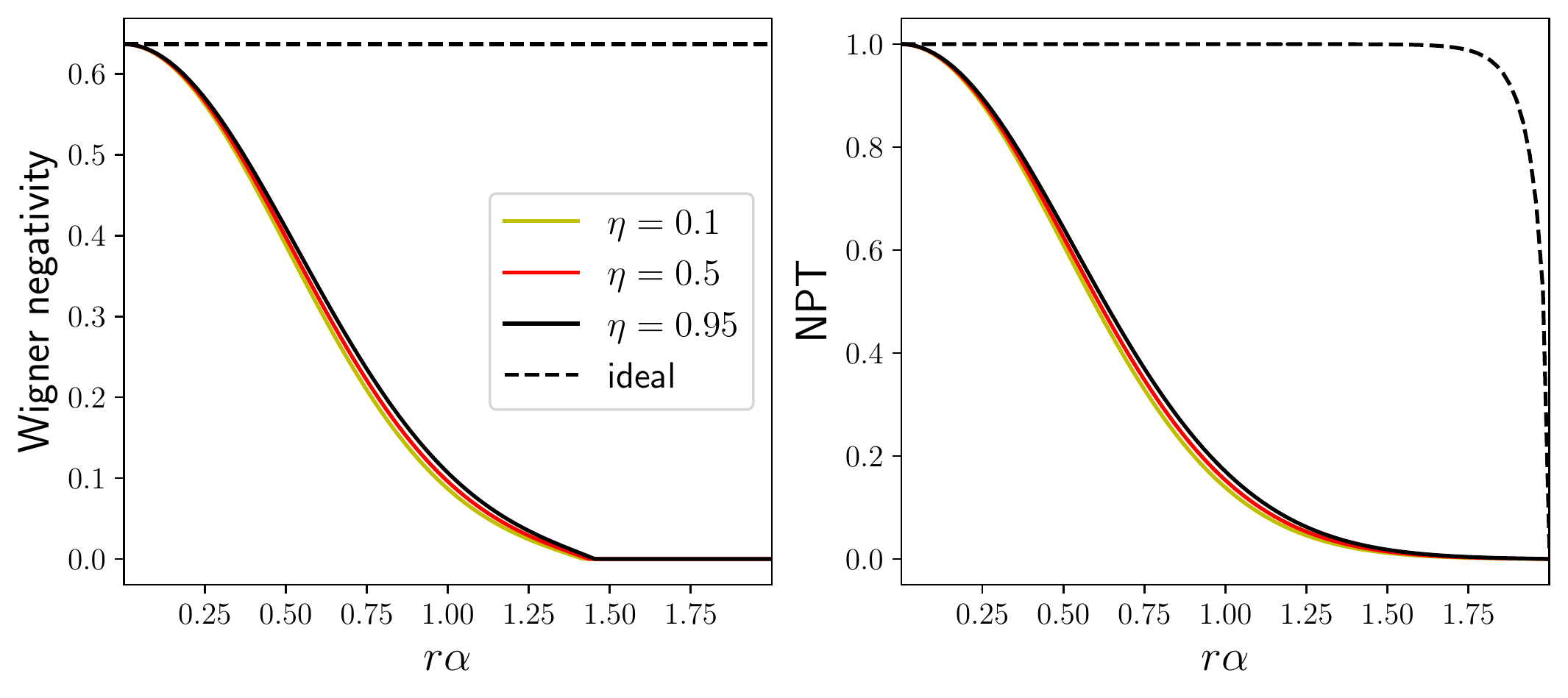}
	\caption{\label{fig:criteres}Negativity of the Wigner function obtained after having projected the DV part of $\hat{\rho}^{(1)}$ onto the state $\frac{\ket{1}_{A, e} + \ket{1}_{A, l}}{\sqrt{2}}$ (left).
	In the ideal case discussed in \autoref{seq:experience_proposee}, the negativity of the Wigner function of the conditional state is constant with $r \alpha$.
	The optimal value of $\frac{2}{\pi} \approx 0.64$ corresponds to the Wigner function definition $W(x, p) = \frac{1}{\pi} \int \mel{x + \frac{u}{2}}{\hat{\rho}^{(1)}}{x - \frac{u}{2}} e^{-2 i p u} \dd{u}$.
	Negativity of the partial transpose (NPT) as a function of $r \alpha$ for the state $\hat{\rho}^{(1)}$ (right).
	The ideal case refers to a hybrid state of \autoref{rhoideal} and to a perfect projective measurement.
	For the nonideal case, quantum efficiencies, $\eta$, of on-off detectors have been chosen according to typical experimental values~\cite{TanzilliJoO2016Quantumphotonicstelecom, ShieldsJoAP2015GigahertzgatedInGaAsInP, NamNP2013Detectingsingleinfrared}.
	Wigner negativity, NPT, and the product $r \alpha$ are adimensional quantities.}
\end{figure}

To conclude, we note that, besides studying the fidelity, in some applied situations, it can be useful to quantify the produced states in terms of specific touchstones~\cite{LoockPRA2012Classifyingquantifyingwitnessing, VogelPRL2017ConditionalHybridNonclassicality}, that can be chosen according to the specific quantum information protocol.
The nonclassicality of the hybrid states as resources for quantum communication is often formalized by their ability to produce a nonclassical continuous variable state when performing a suitable projecting measurement on the discrete variable subsystem.
This operational notion has been theoretically investigated and experimentally validated~\cite{VogelPRL2017ConditionalHybridNonclassicality, LauratPRL2018DemonstrationEinsteinPodolsky}.
In \autoref{fig:criteres}, left, we plot the negativity of the Wigner function of the CV state obtained after having projected the DV part of $\hat{\rho}^{(1)}$ on the state $\frac{\ket{1}_{A, e} + \ket{1}_{A, l}}{\sqrt{2}}$.
Analogous to what is observed for the fidelity, while being weakly dependent on the detection efficiency $\eta$ of the on-off detectors, the negativity of the CV part clearly decreases with $r \alpha$; at the same time, the Wigner function correctly shows negative values for low $r \alpha$ that correspond to the best fidelity with the hybrid target state $\ketbra{\varphi}$.
In order to focus more on the hybrid entanglement quantification, as discussed in~\cite{RalphPS2015Propertieshybridentanglement}, a valuable tool is the negativity of the partial transpose (NPT).
NPT is proportional to the sum of the negative eigenvalues of a partially transposed state density matrix and verifies $0\leq \text{NPT} \leq 1$; separable and Bell states are respectively valued at $0$ and $1$.
The NPT of the state $\hat{\rho}^{(1)}$ is shown by \autoref{fig:criteres}, right.
The state is maximally entangled, NPT is close to $1$, when $r \alpha$ tends to zero, and becomes factorizable when $r \alpha$ increases (NPT is then close to $0$).

As a final remark, we stress that the behaviors of both the Wigner negativity and the NPT can be understood from the expression of $\hat{\rho}^{(1)}$.
In the limit of small $r \alpha$, off-diagonal terms in \autoref{subeq:etat_annonce_on} scale as $-1+\bigO{{(r \alpha)}^2}$ and $\hat{\rho}^{(1)}\sim \ketbra{\phi}+\bigO{{(r \alpha)}^2}$, with no contribution at the first order in $r \alpha$.
This gives the robustness of our state against imperfections and a zero slope to the Wigner negativity and the NPT when $r \alpha\ll 1$.
When $r \alpha$ increases the contribution of off-diagonal terms of the density matrix gently decrease and, accordingly, the purity of announced hybrid state is reduced.
This maps to a decrease of the Wigner negativity and NPT as shown in the figure.

\subsection{Vacuum and multiple pairs in the DV input\label{subseq:vacuum-multi}}
In this section, we further modify our model so as to take into account possible limitations due a more realistic model for the discrete variable input.
The case of a nonideal CV input will be discussed in the last subsection.

So far, we have considered a perfect time-bin entangled state, $\ket{\xi}_{1,2}$, at the input modes $1$ and $2$.
In usual experiments, the generation of time-bin entangled photons typically relies on a parametric down conversion (SPDC) nonlinear process, where a pump photon is converted in a pair of photons (signal and idler)~\cite{ZbindenRoMP2002Quantumcryptography}.
However, this kind of process suffers from unwanted generation of vacuum and multiple pair components~\cite{SilberhornPRA2012Limitsdeterministiccreation}.
In order to discuss their effect, we explicitly include these contributions to the DV input state.
Accordingly, we replace the initial state of \autoref{eq:input_pair} with the state:
\begin{multline}\label{eq:multipair_input}
\ket{\xi'}_{1,2} =
 \sqrt{p_0} \ket{0}_{1,2}
   + \sqrt{p_1} \frac{\ket{1}_{1,e} \ket{1}_{2,e} + \ket{1}_{1,l} \ket{1}_{2,l}}{\sqrt{2}} \\
   + \sqrt{p_\varepsilon} \ket{\varepsilon}_{1,2},
\end{multline}
with $p_0 + p_1 + p_\varepsilon = 1$, and where $\ket{\varepsilon}_{1,2}$ represents all multipair contributions.
A more detailed discussion on the shape of $\ket{\xi'}_{1,2}$ is given in \aref{xi}.
In experimental situations, in order to reduce the impact of multipairs, the SPDC working point is chosen so as to satisfy the condition $p_0 \gg p_1 \gg p_\varepsilon$, thus making the vacuum the most important contribution to $\ket{\xi'}_{1,2}$~\cite{TanzilliLPR2015Ultrafastheralded}.
We will make here this same choice.

As in \autoref{seq:experience_proposee}, we first write the state right before the detection.
This reads
\begin{subequations}
\begin{equation}
\ket{\psi'} = \sqrt{p_0} \ket{\psi^{(0)}}
             + \sqrt{p_1} \ket{\psi^{(1)}}
             + \sqrt{p_\varepsilon}\ket{\psi^{(\varepsilon)}},
\end{equation}
\begin{widetext}
with $\ket{\psi^{(1)}}$ as given by~\eqref{eq:etat_detecteurs},
\begin{multline}\label{subeq:multiple_pairs}
\ket{\psi^{(0)}} =
    \frac{1}{\mathcal{N}} \Big[
      \ket{t \alpha}_B \ket{0}_A
          \hat{D}_{C, l}\left(\frac{r \alpha}{\sqrt{2}}\right) \hat{D}_{D, l}\left(-\frac{r \alpha}{\sqrt{2}}\right)
          \hat{D}_{E, l}\left(-\frac{r \alpha}{\sqrt{2}}\right) \hat{D}_{F, l}\left(\frac{r \alpha}{\sqrt{2}}\right) \ket{0}_{C, D, E, F}\\
      + \ket{-t \alpha}_B \ket{0}_A
          \hat{D}_{C, e}\left(-\frac{r \alpha}{\sqrt{2}}\right) \hat{D}_{D, e}\left(\frac{r \alpha}{\sqrt{2}}\right)
          \hat{D}_{E, e}\left(-\frac{r \alpha}{\sqrt{2}}\right) \hat{D}_{F, e}\left(\frac{r \alpha}{\sqrt{2}}\right)  \ket{0}_{C, D, E, F}\Big],
\end{multline}
\end{widetext}
\end{subequations}
and $\ket{\psi^{(\varepsilon)}}$ a normed state including the contribution due to multiple pairs coming from the discrete part input, $\ket{\xi}_{1,2}$.

As previously, the state is heralded on the simultaneous detection signals $(E, l)$ and $(F, e)$.
As can be seen from~\eqref{subeq:multiple_pairs}, for each of the terms of $\ket{\psi^{(0)}}$, only one of the two temporal modes $(e,l)$ is populated.
Accordingly, $\ket{\psi_{0}}$ has a zero probability to give the heralding signal and it will not contribute to the final heralded state.
Conceptually, the density matrix of the heralded state has the following form:
\begin{align}
\hat{\rho}'
  &= \frac{\Tr_{CDEF}\left[\hat{\mathbf{\Pi}} \ketbra{\psi'} \right]}
          {\Tr[\hat{\mathbf{\Pi}} \ketbra{\psi'}]} \nonumber \\
  &= \frac{1}{1 + \frac{p_\varepsilon P^{(\varepsilon)}}
                       {p_1 P^{(1)}}}
       \left[\hat{\rho}^{(1)}
             + \sqrt{\frac{p_\varepsilon P^{(\varepsilon)}}{p_1 P^{(1)}}}
               \hat{\rho}^{(1,\varepsilon)}
             + \frac{p_\varepsilon P^{(\varepsilon)}}{p_1 P^{(1)}}
               \hat{\rho}^{(\varepsilon)} \right],
	\label{subequ:etat_annonce_vide_et_multi}
\end{align}
with a corresponding heralding probability:
\begin{equation}
P'
  = \Tr[\hat{\mathbf{\Pi}} \ketbra{\psi}]
  = p_1 P^{(1)} + p_\varepsilon P^{(\varepsilon)}.
\end{equation}

In the previous expressions, $P^{(1)}$ is the same as given in \autoref{subeq:taux_annonce_on} and $P^{(\varepsilon)}$ is the probability for $\ket{\psi^{(\varepsilon)}}$ to give a heralding signal.
The explicit expression of $P^{(\varepsilon)}$ can be analytically computed for a given multipair contribution, represented by $\ket{\varepsilon}_{1,2}$.
Similarly, $\hat{\rho}^{(1)}$ is the density matrix already given in \autoref{subeq:etat_annonce_on}, $\hat{\rho}^{(\varepsilon)}$ is a density operator containing multiple photons on mode~$A$, and $\hat{\rho}^{(1,\varepsilon)}$ is a traceless operator whose coefficients are on the order of $1$, as required to preserve the positivity of $\hat{\rho}'$.
For the sake of simplicity, we do not provide the explicit expressions of $\hat{\rho}^{(\varepsilon)}$ and $\hat{\rho}^{(1,\varepsilon)}$ but only discuss their relative weight compared to $\hat{\rho}^{(1)}$.
We note that, as expected, the vacuum component of the discrete variable input is entirely rejected by the heralding process and does not enter the expression of the announced state $\hat{\rho}'$.

The impact of multipair contributions to the heralded state can be quantified in terms of the fidelity of $\hat{\rho}'$ with the target density matrix $\ketbra{\varphi}$.
This reads as
\begin{equation}\label{subeq:fidelite_avec_vide}
\mathcal{F}'
= \frac{1}{1 + \frac{p_\varepsilon P^{(\varepsilon)}}{p_1 P^{(1)}}}
  \mathcal{F}^{(1)},
\end{equation}
with $\mathcal{F}^{(1)}$ as expressed in \autoref{subeq:fidelite_annonce_on}.
For both the announced state, $\hat{\rho}'$, and the fidelity, $\mathcal{F}'$, multiphoton contributions increase with $\frac{p_\varepsilon P^{(\varepsilon)}}{p_1 P^{(1)}}$.
In order to evaluate the general form of this ratio, we note that in most of the situations multipair terms in \autoref{eq:multipair_input} are dominated by double-pair contributions~\cite{TanzilliLPR2015Ultrafastheralded} and, as a consequence, $P^{(\varepsilon)} \approx P^{(2)}$, which is the probability of a double pair to give the heralding signal.
In the limit of a small fraction of light coming from the CV part, it is reasonable to expect $P^{(2)} \underset{r \alpha \to 0}{=} \bigO{\eta^2}$, while, based on \autoref{subeq:taux_annonce_on}, for small $\alpha$, $P^{(1)}\underset{r \alpha \to 0}{=} \eta^2 \abs{r \alpha}^2/32$.
Correspondingly, we obtain
\begin{equation}\label{Ogrand}
\frac{p_\varepsilon P^{(\varepsilon)}}{p_1 P^{(1)}}
	\underset{r \alpha \to 0}{\approx}
		\frac{p_\varepsilon}{p_1} \frac{1}{\bigO{\abs{r \alpha}^2}}.
\end{equation}
In this limit, the fidelity $\mathcal{F}'$ can be written as
\begin{equation}
\mathcal{F}'
  \underset{\mathclap{r \alpha \to 0}}{\approx} \quad
    \frac{1}{1 + \frac{p_\varepsilon}{p_1}
                   \frac{1}{\bigO{\abs{r \alpha}^2}}}
    \left[1 -(1-\frac{\eta}{8}) \abs{r \alpha}^2\right],
\end{equation}
where the numerator is given by~\eqref{subeq:fidelite_annonce_on} for $r \alpha \to 0$.

\hyperref[Ogrand]{Equation\,(\ref{Ogrand})} shows that unwanted multipair contributions to the final heralded state can be avoided provided $p_\varepsilon \ll p_1 \abs{r \alpha}^2$.
In particular, for sources based on second-order nonlinear effects, $p_\varepsilon \sim p_1^2$, and the criterion reads $p_1 \ll \abs{r \alpha}^2$.
By adequately choosing the single pair generation rate $p_1$ and product $r \alpha$ it is thus possible to neglect the components containing multiple photons on the discrete part of the output state.

To conclude with an example, we consider the case of sources based on spontaneous parametric down conversion.
In this case, the explicit expression for $\ket{\xi'}_{1,2}$ can be obtained from the general expression of the output of the SPDC process~\cite{DAuriaPRA2018Quantumdescriptiontiming, LauratPRL2011QuantumDecoherenceSingle}, as a function of an excitation parameter $\lambda^2$ proportional to the pump intensity and to the square of the nonlinear coefficient of the source.
The time-bin state generation can be seen as the result of an SPDC process on mode~$e$ and an SPDC process on mode~$l$.
By combining the coefficients of the two individual processes, we obtain for the overall generation the weights (see \aref{xi})
\begin{equation}\label{pSPDC}
\begin{split}
	p_0 &= {\left(1-\lambda^2\right)}^2,
	\\
	p_1 &= 2 {\left(1-\lambda^2\right)}^2 \lambda^2,
	\\
	p_2 &= 3 {\left(1-\lambda^2\right)}^2 \lambda^4.
\end{split}
\end{equation}
The explicit expression for the $P^{(2)}$ is
\begin{small}
\begin{equation}\label{P2}
P^{(2)} = \frac{\eta}{48}
	\frac{12 - \eta - (12-2\eta+\frac{\eta^2}{2}\abs{r\alpha}^2)
	e^{-\frac{\eta}{2}\abs{r\alpha}^2}
	+\eta e^{-2 \abs{\alpha}^2}}{1+e^{-2 \abs{\alpha}^2}}.
\end{equation}
\end{small}
In the limit of $r \alpha \to 0$, this expression leads to $P^{(2)}\underset{r \alpha \to 0}{=} \eta^2/48$ and, as expected, correctly behaves as $\eta^2$.
More explicitly, based on Eqs.\,\eqref{pSPDC}~and~\eqref{P2}, we find for the SPDC
\begin{equation}\label{LimitSPDC}
\frac{p_2 P^{(2)}}{p_1 P^{(1)}} \underset{r \alpha \to 0}{\approx} \frac{\lambda^2}{\abs{r \alpha}^2}.
\end{equation}
The \autoref{fig:fidelite_vs_p1_on} (dashed line) shows the fidelity $\mathcal{F}'$ as a function of the parameter $\lambda^2$.
We considered, as for the previous section, $\eta = 0.95$, $\frac{r \alpha}{\sqrt{2}} = 0.075$, with $\alpha=2$ and $r = 0.05$.
As seen, for these values, the fidelity in absence of multipair contributions is $\mathcal{F}\approx0.99$.
The fidelity $\mathcal{F}'$ approaches $\mathcal{F}$ when $\lambda^2 \ll \abs{r \alpha}^2$ (see \autoref{LimitSPDC}), \latin{i.e.\@} $\lambda^2 \ll 0.01$, and decreases when the contribution of multipair increases.

\subsection{Squeezed states at the CV input}
In this last section, we briefly discuss the case of nonideal states on the CV input.
So far, we have considered at the CV input a perfect Schrödinger cat state as described by \autoref{cat}.
Nevertheless, these states are difficult to generate experimentally and they are often replaced by Schrödinger kitten states, of small size $\alpha$, generated in a heralded fashion~\cite{LvovskyNP2017EnlargementopticalSchroedingers,GrangierN2007GenerationopticalSchroedinger}.
At the same time, the use of such heralded states as a starting resource for the generation of hybrid-entanglement implies a further heralding signal to be combined with the ones considered so far (\latin{i.e.\@} $(E,l)$ and $(F,e)$).
This would imply that the overall protocol success would rely on a threefold coincidence signal with a dramatic effect for the generation rate.

\begin{figure}[h]
	\centering
	\includegraphics[width=\linewidth]{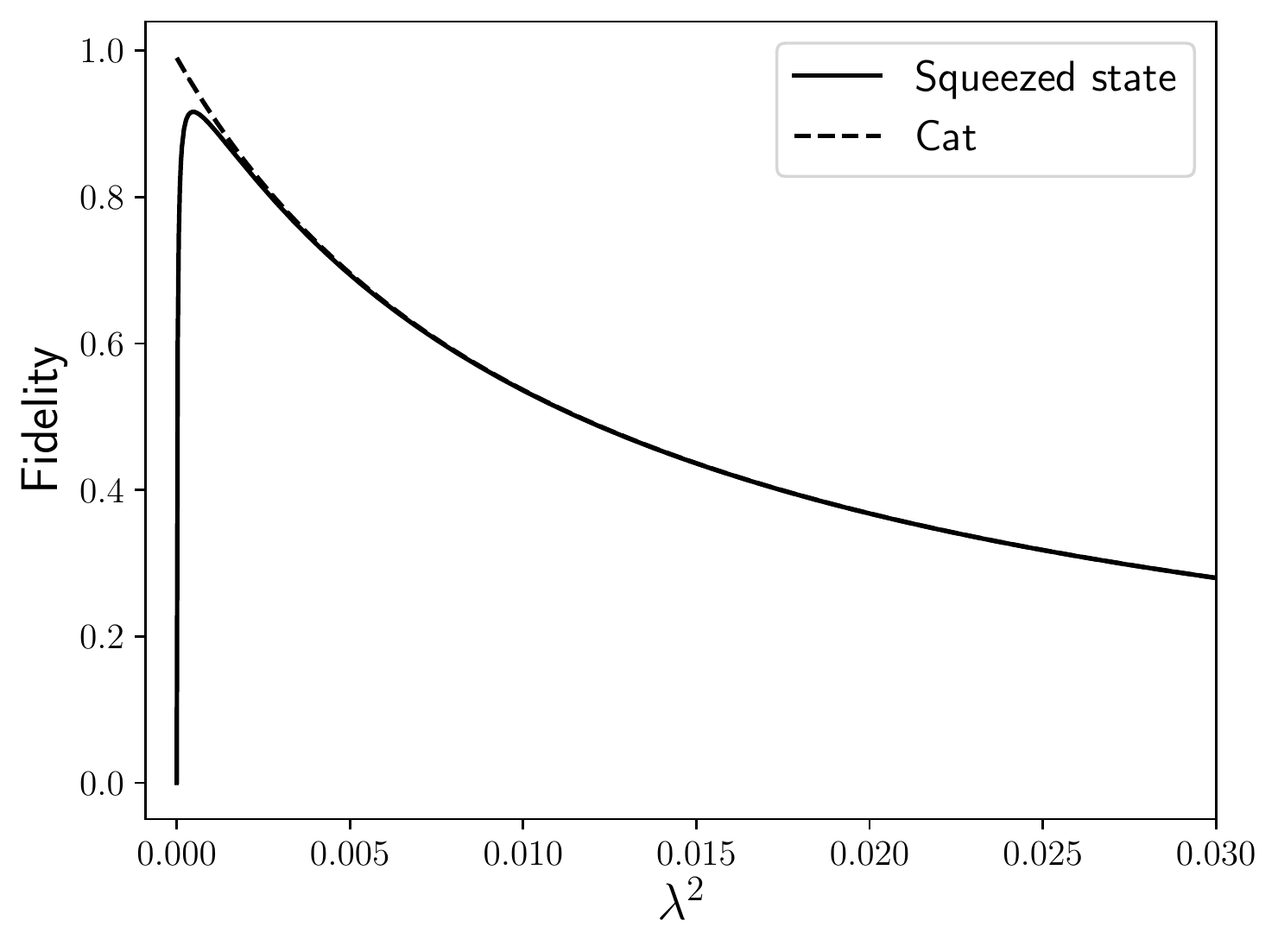}
	\caption{\label{fig:fidelite_vs_p1_on}
		Fidelities with respect to $\ket{\varphi}$ of the states obtained when considering multiple pairs on the discrete variable input as functions of the excitation parameter $\lambda^2$ of a realistic source based on parametric down conversion and providing the DV input state.
		Cat and squeezed states were considered for the continuous variable input.
		Detection efficiency $\eta = 0.95$.
		We chose $\frac{r \alpha}{\sqrt{2}} = 0.075$, with a value of $\alpha=0.25$.
		Fidelity and $\lambda^2$ are adimensional quantities.}
\end{figure}

In view of realistic realizations, a convenient approximation to the state $\ket{\text{cat+}}$ with $\alpha<1$ consists in using at the input $3$ a squeezed vacuum state~\cite{LvovskyNC2018Entanglementteleportationpolarization}, $\hat{S}(\zeta)\ket{0}_3$, where $\hat{S}(\zeta)=e^{\frac{1}{2}(\zeta^* \hat{a}^2-\zeta\hat{a}^{\dagger 2})}$ is the single mode squeezing operator and $\zeta$ is the squeezing parameter.
These states are deterministically available at the output of many nonlinear optical systems with a huge simplification of required experimental resources~\cite{LeuchsPS201630yearssqueezed}.
The main drawback of this approach is that the interference of light reflected from BS1 with the coherent input $\ket{r \alpha}$ at BS2 is no longer perfect, thus modifying the shape of the state of \autoref{eq:after_BS2} and in turns of $\ket{\psi^{(1)}}$.
Under these conditions, the heralding signals $(E,l)$, $(F,e)$ could in principle be triggered from the sole CV part and occurring even in the absence of photons in the discrete variable.
This would lead to an announced state containing vacuum contributions in its DV part.

The described events are linked to vacuum contributions in the DV input $\ket{\xi'}_{1,2}$.
Accordingly, in analogy to the formalism of the previous paragraph, we will call $P^{(0)}$ their corresponding probability and write the overall heralding probability as $P'_s=p_0P^{(0)}+p_1P^{(1)}+p_\epsilon P^{(\epsilon)}$, with $P^{(1)}$ and $P^{(\epsilon)}$ already discussed in the previous sections.
We note that in realistic situations, to comply with multipair events, the nonlinear process providing the DV input $\ket{\xi'}_{1,2}$ is weakly pumped.
This condition comes at the price of a high vacuum contribution represented by a $p_0$ close to $1$.
By taking into account both the effects of vacuum and multipairs due to nonideal DV and CV inputs, we can express the fidelity as
\begin{equation}\label{subeq:fidelite_avec_vide_squeeze}
\mathcal{F}'_s
= \frac{1}{1 + \frac{p_0 P^{(0)}}{p_1 P^{(1)}}
           + \frac{p_\varepsilon P^{(\varepsilon)}}{p_1 P^{(1)}}}
  \mathcal{F}^{(1)},
\end{equation}
where we have neglected minor changes on $P^{(1)}$ and $\mathcal{F}^{(1)}$.
The possibility of having a heralding signal with vacuum on the DV input further reduces the fidelity with respect to the situation described in \autoref{subseq:vacuum-multi}.
To avoid the effect of vacuum contribution, $\frac{p_0 P^{(0)}}{p_1 P^{(1)}}\ll1$.
As for \autoref{subseq:vacuum-multi}, it is pertinent to analyze this ratio in the limit $r \alpha \to 0$, where $ \mathcal{F}^{(1)}$ is close to 1.
The analytical expression of $P^{(0)}$ is reported in \aref{P0}.
It depends on the input squeezing level, $\zeta$, on the product $\abs{r\alpha}$, and on the detection efficiency $\eta$.
Its asymptotic behavior gives $P^{(0)} \underset{r \alpha \to 0}{=} \bigO{\eta^2 {\left(\frac{\abs{r \alpha}^2}{r}\right)}^4}$.
Accordingly, by considering $p_0 = \bigO1$ and by taking the limit of \autoref{subeq:taux_annonce_on} for $P^{(1)}$, we obtain
\begin{equation}\label{Ogrand0}
\frac{p_0 P^{(0)}}{p_1 P^{(1)}} \underset{r \alpha \to 0}{\propto}
\frac{\abs{r \alpha}^6}{p_1 {r}^4}.  % facteur numérique : 32
\end{equation}
By putting together Eqs.\,\eqref{Ogrand}~and~\eqref{Ogrand0}, we obtain that, with realistic input states and detectors, optimal fidelity $\mathcal{F}'_s$ is obtained when
\begin{equation}\label{optimalp1}
\frac{\abs{r \alpha}^6}{{r}^4}\ll p_1 \ll \abs{r \alpha}^2.
\end{equation}
Compared to the case of an ideal input cat state $\ket{\text{cat+}}$, where the fidelity is maximized by taking an arbitrary low value of $p_1$ below $ \abs{r \alpha}^2$, when considering at the CV input a squeezed state, an optimal value of $p_1$ must be chosen so as to comply with both conditions of \autoref{optimalp1}.
Better approximation of Schrödinger cat states than squeezed states~\cite{LvovskyNP2017EnlargementopticalSchroedingers,GrangierN2007GenerationopticalSchroedinger} would even reduce the lower bound of the previous inequality, thus allowing one to reach a better fidelity with respect to $\ket{\varphi}$.

To conclude, in \autoref{fig:fidelite_vs_p1_on} we illustrate the behavior of the fidelity $\mathcal{F}'_s$ in the case of a DV input state $\ket{\xi'}_{1,2}$ produced by SPDC as given by \hyperref[pSPDC]{Eqs.\,(\ref{pSPDC})}.
The fidelity is reported as a function of the SPDC excitation parameter, $\lambda^2$, and it is compared with the $\mathcal{F}'$ corresponding to an ideal CV input state, $\ket{\text{cat+}}_{3}$ (dashed line).
In the numerical computation, we have considered, as for the previous paragraph, $\eta = 0.95$ and $\frac{r \alpha}{\sqrt{2}} = 0.075$.
In particular, we have imposed for the CV input state the squeezing level $\zeta$ minimizing $P^{(0)}$ under the condition $\frac{r \alpha}{\sqrt{2}} = 0.075$ and by choosing $\alpha=0.25$.
In these regards, we note that, in experiments, low $\alpha$ values allow one to satisfy the approximation of a Schrödinger kitten state with a squeezed state~\cite{LauratNP2014Remotecreationhybrid}.
With this set of values, the minimum $P^{(0)} = 1.3 \times {10}^{-8}$ is obtained for a squeezing parameter $\zeta = -0.061$ and leads to the fidelity given in \autoref{fig:fidelite_vs_p1_on}.
The optimal value, $\mathcal{F}'_s = 0.92$, corresponds to $\lambda^2 = 9.4 \times 10^{-4}$ % Lvovsky: $0.83$, Julien: $0.77$. \
and to a heralding probability of $P'_s \approx 3\times {10}^{-7}$.
We note that this value is consistent with similar estimated~\cite{JeongPRA2015Generationhybridentanglement, WangPRA2018Experimentallyfeasiblegeneration} and measured~\cite{LvovskyPRL2017QuantumTeleportationDiscrete} values in schemes where hybrid state generation is conditioned upon a double detection signal.
By operating the experiment at \unit{1}{\giga\hertz} repetition rate~\cite{TanzilliLPR2015Ultrafastheralded}, the heralding rate reaches \unit{300}{\hertz}.
If needed, a higher heralding probability can be reached, at the cost of lower values of the fidelity.
We note that compromises between the heralding rate and the quality of the produced states are also required in the generation of other kinds of hybrid DV-CV entanglement~\cite{LvovskyNC2018Entanglementteleportationpolarization, LauratNP2014Remotecreationhybrid} and, more generally, in any heralded state preparation.

\section{Conclusion}

In this article, we have presented an experimental scheme able to generate a time-bin encoded hybrid entangled state of the form $\ket{\varphi} = \frac{\ket{1}_{A,e}\ket{+\alpha_f}_B - \ket{1}_{A,l}\ket{-\alpha_f}_B}{\sqrt{2}}$.
Our protocol is fully compatible with experimental realizations with off-the-shelf fiber components.
The required input resources are a coherent state, an optical Schrödinger cat state $\ket{\text{cat+}}$ and a time-bin entangled photon pair $\ket{\xi}_{1,2}$.
We have shown that, with ideal perfect photon-number-resolving detectors, the process exactly generates the desired state $\ket{\varphi}$.
In the second part of the paper, we studied the case of nonideal detectors and/or input states.
In particular, we have shown that even with realistic detectors, \latin{i.e.\@} available from today's commercial technology, an arbitrary close approximation of $\ket{\varphi}$ can be obtained by playing on the scheme parameters.
A major advantage of the presented scheme lies in its ability to tolerate both predominance of vacuum on its discrete variable input, as well as multiple pairs, when operated with well-chosen parameters.
In this context, we have studied the case of a realistic DV input as well as the one of a CV one in a squeezed vacuum state instead of the Schrödinger cat state.
Our study shows that vacuum and multipair effects can be neglected as long as the experiment is carried out with a $p_1$ in the DV input respecting the condition $\frac{\abs{r \alpha}^6}{{r}^4} \ll p_1 \ll \abs{r \alpha}^2$.

Experimental generation of time-bin coded hybrid states, compatible with standard telecommunication systems, will permit pushing the applications of hybrid states of light out of the laboratory with a high impact in the context of future development of fiber quantum network systems.

\begin{acknowledgments}
All the authors would like to thank Jean Etesse and Anthony Martin for stimulating discussions.
This work has been conducted within the framework of the project HyLight (No.~ANR-17-CE30-0006-01) granted by the Agence Nationale de la Recherche (ANR).
The authors also acknowledge financial support from the European Union and the Région PACA by means of the Fond Européen de Développement Regional (FEDER) through the project OPTIMAL and the French government through its program ``Investments for the Future'' under the Université Côte d'Azur UCA--JEDI project (under the label Quantum@UCA) managed by the ANR (Grant agreement No.~ANR-15-IDEX-01).
\end{acknowledgments}

\appendix

\section{Action of losses}\label{Loss}

We consider losses on DV and the CV parts of state $\ket{\varphi}$ defined by \autoref{eq:output_state}.
Similar expressions can be found for the states obtained in nonideal conditions and discussed in different sections of the paper.
Following a very standard approach, losses are modeled as unbalanced beam splitters inserted along the paths of mode~$A$ and mode~$B$ of the state.
The higher the losses, the higher are the reflection coefficients.
The other BS input of each of these fictitious BS is in a vacuum state.

\subsection{Losses on hybrid states with time-bin encoding}
Losses on the CV part of $\ket{\varphi}$ are modeled by an unbalanced beam splitter of transmission coefficient $t_{\text{CV}}$ and reflection coefficient $r_{\text{CV}}$ that couple the CV channel $B$ with a vacuum populated channel.
We will label as $B'$ the BS transmitted output and as ``$\text{lost}$'' the reflected one, that also corresponds to the lost part.
Hence the state after the beam splitter is
\begin{multline}
\ket{\varphi'} = \frac{1}{\sqrt{2}}\Big[\ket{1}_{A,e}\ket{+ r_{\text{CV}}\alpha_f}_{\text{lost}}\ket{+ t_{\text{CV}}\alpha_f}_{B'}\\[-1ex]
- \ket{1}_{A,l}\ket{- r_{\text{CV}}\alpha_f}_{\text{lost}}\ket{- t_{\text{CV}}\alpha_f}_{B'}\Big].
\end{multline}
The density matrix describing the state after losses on the mode~$B$ is obtained by tracing out on the mode~``$\text{lost}$'':
\begin{equation}\label{rhocvloss}
\begin{split}
\hat{\rho}' &= \text{Tr}_{\text{lost}}\left[\ketbra{\varphi'}\right] \\
& = \frac{1}{2}\biggl[\ketbra{1}{1}_{A,e}\ketbra{+ t_{\text{CV}}\alpha_f}{+ t_{\text{CV}}\alpha_f}_{B'}\\
	&\quad - e^{-2\abs{r_{\text{CV}}\alpha_f}^2}(\ketbra{1}{0}_{A,e}\ketbra{0}{1}_{A,l} \ketbra{+ t_{\text{CV}}\alpha_f}{- t_{\text{CV}}\alpha_f}_{B'}\\
	&\quad + \ketbra{0}{1}_{A,e}\ketbra{1}{0}_{A,l} \ketbra{- t_{\text{CV}}\alpha_f}{+ t_{\text{CV}}\alpha_f}_{B'})\\
	&\quad + \ketbra{1}{1}_{A,l}\ketbra{- t_{\text{CV}}\alpha_f}{- t_{\text{CV}}\alpha_f}_{B'}\biggr].
\end{split}
\end{equation}
When losses on the CV part increase, the coefficient $e^{-2\abs{r_B\alpha_f}^2}$ multiplying the off-diagonal terms of $\hat{\rho}'$ decreases, thus resulting in a decoherence effect.
Although we report here the calculation for an initial state of $\ketbra{\varphi}$, this effect is common to any hybrid CV-DV entangled state submitted to losses in the CV part.

Losses on the DV part are modeled by an unbalanced beam splitter of transmission coefficient $t_{\text{DV}}$ and reflection coefficient $r_{\text{DV}}$ that couple the DV channel~$A$ with a vacuum populated channel.
We will label as $A'$ the BS transmitted output and as ``$\text{lost}$'' the reflected one, that also corresponds to the lost part.
Hence the state after the beam splitter is
\begin{multline}\label{LossDVTB}
\ket{\tilde{\varphi}} = \frac{1}{\sqrt{2}} \Big[
	t_{\text{DV}} \ket{1}_{A', e}\ket{+\alpha_{f}}_{B}
		+ r_{\text{DV}} \ket{1}_{\text{lost}, e}\ket{+\alpha_{f}}_{B} \\[-1ex]
	- t_{\text{DV}}\ket{1}_{A', l}\ket{-\alpha_{f}}_{B}
		- r_{\text{DV}} \ket{1}_{\text{lost}, l}\ket{-\alpha_{f}}_{B}
\Big].
\end{multline}
By tracing out the modes $(\text{lost}, e)$ and $(\text{lost}, l)$, we obtain the state after losses, described by the density matrix:
\begin{equation}
\begin{split}
\tilde{\rho}
	&= \Tr_{\text{lost}}\left[\ketbra{\tilde{\varphi}}\right] \\
	&= \begin{multlined}[t]
	t_{\text{DV}}^2 \ketbra{\varphi} \\ + r_{\text{DV}}^2 \ketbra{0}_{A'}
		\frac{\ketbra{+\alpha_{f}}_{B} + \ketbra{-\alpha_{f}}_{B}}{2}.
	\end{multlined}
\end{split}
\end{equation}
It is clear that, for any projection operation (or measurement) on $A'$ that is insensitive to vacuum, the result is the same as the one obtained from $\ketbra{\varphi}$, with a success probability multiplied by $t_\text{DV}^2$.

Upon the combined effect of CV and DV losses, it is easy to show that the density matrix of the initial state $\ketbra{\varphi}$ becomes
\begin{multline}
\hat{\rho}_{\text{loss}}
	= t_{\text{DV}}^2 \hat{\rho}' \\[-1ex]
	+ \frac{r_{\text{DV}}^2}{2} \ketbra{0}_{A'}
		\Big[\ketbra{+t_{\text{CV}}\alpha_{f}}_{B} +\\[-1ex] \ketbra{-t_{\text{CV}}\alpha_{f}}_{B}\Big].
	\end{multline}
with $\hat{\rho}' $ given by \autoref{rhocvloss}.

\subsection{Propagation effect on hybrid states with polarization encoding}
The state considered here is a hybrid state with the DV part encoded on polarization~\cite{WangPRA2018Experimentallyfeasiblegeneration, LvovskyNC2018Entanglementteleportationpolarization}:
\begin{equation}
\ket{\varphi_{\text{pol}}} = \frac{\ket{1}_{A,H}\ket{+\alpha}_B  - \ket{1}_{A,V}\ket{-\alpha}_B}{\sqrt{2}},
\end{equation}
where $H$ stands for horizontal and $V$ for vertical polarization.
For this DV encoding, the effect of losses on the CV and DV part is the same as for time-bin DV encoding.
However, polarization encoded photons strongly suffer from polarization variations when they propagate in long optical fibers~\cite{UrsinPotNAoS2019Entanglementdistributionover}.
To describe the loss of information about the polarization during the propagation, we follow the approach of Ref.\,\cite{MenyakJoLT1996Polarizationmodedispersion} and divide the fiber into small sections of length $z$ whose local birefringence rotates light polarization of an angle $\theta$.
The state after a rotation is expressed as
\begin{multline}\label{rhodepol}
\ket{\varphi_{\text{pol},\theta}}= \frac{1}{\sqrt{2}}\Big[\Big(\cos(\theta)\ket{1}_{A,H} - \sin{\theta}\ket{1}_{A,V}\Big)\ket{+\alpha}_B\\
 - \Big(\sin(\theta)\ket{1}_{A,H} +\cos{\theta}\ket{1}_{A,V}\Big)\ket{- \alpha}_B\Big].
\end{multline}
The density matrix of state at the output of the fiber is then obtained by averaging on the angle $\theta$ the $\ket{\varphi_{\text{pol},\theta}}$ weighted by the angle distribution:
\begin{equation}
\hat{\rho}^{\text{depol}}(z) = \int \operatorname{P}(\theta, z)\ketbra{\varphi_{\text{pol},\theta}} \dd{\theta},
\end{equation}
where
$\operatorname{P}(\theta, z) = \frac{1}{\sqrt{2\pi\sigma^2z}}e^{-\frac{\theta^2}{2\sigma^2z}}$ is a Brownian probability density subjected to impulsive changes as the fiber length, $z$, increases.
The parameter $\sigma$ is defined as $\sigma=\sqrt{\frac{2}{L_C}}$, with $L_c$ the length over which the angles $\theta$ lose correlation~\cite{MenyakJoLT1996Polarizationmodedispersion}.
The expression~\eqref{rhodepol} gives
\begin{multline}\label{eq:densite_pola1}
\hat{\rho}^{\text{depol}}(z) = \frac{1}{2} (1 + e^{-2\sigma^2z}) \ketbra{\varphi}\\+ \frac{1}{2} (1 - e^{-2\sigma^2z}) \ketbra{\varphi_{\text{pol},-\pi/2}},
\end{multline}
where we omitted the spatial mode labels $A'$ and $B$.
As seen from the previous expression, polarization dispersion can progressively convert vertical polarization, \latin{i.e.\@} DV state, $\ket{1}_{A,V}$, into the horizontal one $\ket{1}_{A,H}$ and vice versa, thus leading to a deterioration of the state purity with the distance.
For very long distances, the state $\hat{\rho}^{\text{depol}}(z)$ is a mixture of $\ket{\varphi_{\text{pol}}}$ and the state $\ket{\varphi_{\text{pol},-\pi/2}}$ is obtained when exchanging the $H$ and $V$ in $\ket{\varphi_{\text{pol}}}$.

\subsection{Propagation effect on hybrid states with single rail encoding}
We consider here a hybrid entangled state on which the DV part is encoded on the presence $\ket{1}_A$ and the absence $\ket{0}_A$ of a single photon in mode~$A$, \latin{i.e.\@} exhibiting single-rail DV encoding~\cite{BelliniNP2014Generationhybridentanglement,
LauratNP2014Remotecreationhybrid}:
\begin{equation}
\ket{\varphi_{\text{s-r}}} = \frac{\ket{1}_A\ket{+\alpha_f}_B  - \ket{0}_A\ket{-\alpha_f}_B}{\sqrt{2}}.
\end{equation}
The state after the loss beam splitter is expressed as
\begin{multline}
\ket{\tilde{\varphi}_{\text{s-r}}} = \frac{1}{\sqrt{2}} \biggl( ( t_\text{DV}\ket{1}_{A'}\ket{0}_{\text{lost}} + r_\text{DV}\ket{0}_{A'}\ket{1}_{lost})\ket{+\alpha_f}_B \\
 - \ket{0}_{A'}\ket{0}_{lost}\ket{- \alpha_f}_B \biggr).
\end{multline}
where as before we label as $A'$ the BS transmitted output and as ``$\text{lost}$'' the reflected one.
By tracing out on the mode~``$\text{lost}$'', we obtain
\begin{equation}\label{lossSingleDV}
\begin{split}
\tilde{\rho}_{\text{s-r}}  = \frac{1}{2} \biggl(&(t_\text{DV}^2\ketbra{1}{1}_{A'} + r_\text{DV}^2\ketbra{0}{0}_{A'})\ketbra{+\alpha}{+\alpha}_B\\
& - t_\text{DV}\ketbra{1}{0}_{A'} \ketbra{+\alpha}{-\alpha}_B\\
& - t_\text{DV}\ketbra{0}{1}_{A'} \ketbra{-\alpha}{+\alpha}_B\\
& + \ketbra{0}{0}_{A'}\ketbra{-\alpha}{-\alpha}_B\biggr).
\end{split}
\end{equation}
The coefficient $t_\text{DV}$ is related to the fiber length by the Beer Lambert law: $t_\text{DV} = e^{-\frac{1}{2}\beta z}$ where the coefficient $\beta$ is the linear absorption coefficient of the fiber at the working wavelength.
We observe that, for single-rail encoding, losses change the relative weight of the DV qubit term by reducing the contribution of $\ket{1}$, in favor of $\ket{0}$.
This effect cannot be eliminated by postselection at the detection stage and degrades the hybrid entanglement.

\subsection{Remote preparation of CV qubit}\label{LossRemote}

We consider a remote CV qubit preparation experiment~\cite{LauratO2018Remotepreparationcontinuous} and compare the performances of three kinds of hybrid entangled states with different DV encodings and in the case of a DV part traveling over long distances in optical fibers.

We start with the case of a hybrid entangled state with time-bin encoding submitted to loss on its DV part as in \autoref{LossDVTB}.
With no loss of generality, we consider the case of a DV measurement described by the projector on $\frac{\ket{1}_{A', e} + \ket{1}_{A', l}}{\sqrt{2}}$ leading to an odd Schrödinger cat state on the CV part of the state.
The associated measurement operator is
\begin{equation}
\hat{\Pi}_{A'}
	= \frac{1}{2} \left(\ket{1}_{A', e} + \ket{1}_{A', l}\right)
		\left(\bra{1}_{A', e} + \bra{1}_{A', l}\right).
\end{equation}
The unnormalized state on the CV part after such a conditioning reads
\begin{multline}
\Tr_{A'}[\hat{\Pi}_{A'} \hat{\rho}'] \\
\qquad\begin{aligned}
	&= \frac{t_{\text{DV}}^2}{4} \left[\ket{+\alpha_{f}}_{B} - \ket{-\alpha_{f}}_{B}\right]
		\left[\bra{+\alpha_{f}}_{B} - \bra{-\alpha_{f}}_{B}\right] \\
	&\propto \ketbra{\text{cat-}}_{B}.
\end{aligned}
\end{multline}
The obtained state is exactly the same as it would be without losses and it has a unit fidelity with the target, whatever the propagation distance.

A similar analysis can be done on remote preparation of an odd Schrödinger cat state starting with hybrid entanglement with polarization encoding after the propagation of the DV part as in \autoref{eq:densite_pola1}.
In this case, the measurement on the DV part of the state $\hat{\rho}^{\text{depol}}$ is described by the projector $\hat{\Pi}_A = \frac{1}{2}(\ket{1}_{H,A} + \ket{1}_{V,A})(\bra{1}_{H,A} + \bra{1}_{V,A})$.
The fidelity of the so obtained CV state and an odd cat state $\ket{\text{cat}_-}$ is expressed as
\begin{equation}
\mathcal{F}_{\text{pol}}(z) = \frac{1}{2} + \frac{e^{-\frac{z}{L_C}} - e^{-2\abs{\alpha_f}^2}}{2(1 - e^{-\frac{z}{L_C}}e^{-2\abs{\alpha_f}^2})}.
\end{equation}
Eventually, by starting from the state of \autoref{lossSingleDV} and following the same approach as for remote state preparation with hybrid states with time-bin or polarization DV encoding, the fidelity with the target odd Schrödinger cat state is
\begin{equation}
\mathcal{F}_{\text{s-r}}(z) = \frac{1}{2} + \frac{e^{-\frac{1}{2}\beta z} - e^{-2|\alpha|^2}}{2(1 - e^{-\frac{1}{2}\beta z}e^{-2|\alpha|^2})}.
\end{equation}
We observe that for both polarization and single-rail DV encodings, the fidelity of the obtained state with the target one is degraded when long propagation distances, $z$, are considered.

\section{Time-bin entangled photons source}\label{xi}
\paragraph*{Generic form}
Typically, the time-bin entangled photon pair $\ket{\xi'}_{1,2}$ (see \autoref{eq:multipair_input}) can be seen as the result of two identical entangled photon pair generation processes, one for the early ($e$) and other for the late ($l$) mode.
In the Fock basis, whose elements are indicated here as $\{\ket{k}\}$, the generated state can be written as
\begin{widetext}
\begin{multline}\label{doppiasource}
\ket{\xi'}_{1,2}
	= \left(\sqrt{p_0^\text{m}} \ket{0}_{1,e} \ket{0}_{2,e}
			+ \sqrt{p_1^\text{m}} \ket{1}_{1,e} \ket{1}_{2,e}
			+ \sqrt{p_2^\text{m}} \ket{2}_{1,e} \ket{2}_{2,e}
			+ \ldots\right) \\
		\otimes
		\left(\sqrt{p_0^\text{m}} \ket{0}_{1,l} \ket{0}_{2,l}
			+ \sqrt{p_1^\text{m}} \ket{1}_{1,l} \ket{1}_{2,l}
			+ \sqrt{p_2^\text{m}} \ket{2}_{1,l} \ket{2}_{2,l}
			+ \ldots\right).
\end{multline}

By explicitly taking into account only terms containing at most two photons per spatial mode, we can write
\begin{equation}
\ket{\xi'}_{1,2}
	\approx p_0^\text{m} \ket{0}
		+ \sqrt{2 p_0^\text{m} p_1^\text{m}}
			\frac{\ket{1}_{1,e}\ket{1}_{2,e}+ \ket{1}_{1,l}\ket{1}_{2,l}}{\sqrt{2}}
		+ \sqrt{2 p_0^\text{m} p_2^\text{m}}
			\frac{\ket{2}_{1,e}\ket{2}_{2,e}
		+ \ket{2}_{1,l}\ket{2}_{2,l}}{\sqrt{2}}
		+ p_1^m \ket{1}_{1,e}\ket{1}_{1,l} \ket{1}_{2,e}\ket{1}_{2,l},
\end{equation}
where, as before, $\ket{1}$ and $\ket{2}$ are single and two photon Fock states, respectively.
\end{widetext}

By comparing the previous expression with the general form of $\ket{\xi'}_{1,2}$ cut at the second order ($p_\varepsilon \approx p_2$), we obtain the values of coefficients appearing in \autoref{eq:multipair_input}:
\begin{equation}\label{eq:generic_two_spdc}
\begin{split}
	p_0 &= {(p_0^m)}^2,
	\\
	p_1 &= 2 p_0^\text{m} p_1^\text{m},
	\\
	p_2 &= 2 p_0^\text{m} p_2^\text{m} + {(p_1^\text{m})}^2.
\end{split}
\end{equation}
with the second-order term being
\begin{multline}
\ket{\epsilon}_{1,2} =
	\sqrt{\frac{2 p_0^\text{m} p_2^\text{m}}
			{2 p_0^\text{m} p_2^\text{m} + {(p_1^\text{m})}^2}}
		\frac{\ket{2}_{1,e}\ket{2}_{2,e}
			+ \ket{2}_{1,l}\ket{2}_{2,l}}{\sqrt{2}}\\
	+ \frac{p_1^\text{m}}
			{\sqrt{2 p_0^\text{m} p_2^\text{m} + {(p_1^\text{m})}^2}}
		\ket{1}_{1,e}\ket{1}_{1,l} \ket{1}_{2,e}\ket{1}_{2,l}.
\end{multline}

\paragraph*{SPDC case}
The explicit expression of coefficients $p_k^\text{m}$ ($k=0,1,2\ldots$) in the previous equations depends on the specific generation process.
In the special case of SPDC, $p_k^\text{m} = \left(1-\lambda^2\right) {(\lambda^2)}^k$~\cite{DAuriaPRA2018Quantumdescriptiontiming, LauratPRL2011QuantumDecoherenceSingle}, with $\lambda$ the SPDC excitation parameter as described in the text.
By injecting these expressions in \autoref{eq:generic_two_spdc}, we obtain the results of \autoref{pSPDC}:
\begin{equation}
\begin{split}
	p_0 &= {\left(1-\lambda^2\right)}^2,
	\\
	p_1 &= 2 {\left(1-\lambda^2\right)}^2 \lambda^2,
	\\
	p_2 &= 3 {\left(1-\lambda^2\right)}^2 \lambda^4.
\end{split}
\end{equation}
We also observe that, for SPDC, the two photon component reads
\begin{align}\label{eq:poids_relatif_source_dv}
\ket{\epsilon}_{1,2} &=
	\begin{multlined}[t][20em]
\sqrt{\frac{2}{3}}
	\frac{\ket{2}_{1,e}\ket{2}_{2,e} + \ket{2}_{1,l}\ket{2}_{2,l}}{\sqrt{2}} \\[-3ex] + \sqrt{\frac{1}{3}} \ket{1}_{1,e}\ket{1}_{1,l} \ket{1}_{2,e}\ket{1}_{2,l}
	\end{multlined} \\ \notag
& = \frac{\ket{2}_{1,e}\ket{2}_{2,e} + \ket{2}_{1,l}\ket{2}_{2,l} + \ket{1}_{1,e}\ket{1}_{1,l} \ket{1}_{2,e}\ket{1}_{2,l}}{\sqrt{3}}.
\end{align}

%\bigskip  % Artificiel pour forcer LaTeX à être gentil.

\section{Heralding probability with a CV input squeezed vacuum state and with no photon from the DV input}\label{P0}
We consider at the CV a squeezed vacuum state, $\hat{S}(\zeta)\ket{0}_3$, with $\hat{S}(\zeta)=e^{\frac{1}{2}(\zeta^* \hat{a}^2-\zeta\hat{a}^{\dagger 2})}$ the single mode squeezing operator and $\zeta$ the squeezing parameter.
In this case, the probability of having an announced signal with no photon from the DV part is not zero and it is a function of $\zeta$, of the detection efficiency $\eta$, and of the amplitude $r \alpha $ of the coherent state entering the system via the input labeled as $4$.
As for the previous cases, $r$ and $t$ are the amplitude reflection and transmission coefficients of BS1:
\begin{widetext}
\begin{equation}
P_0 = \frac{e^{-\abs{r \alpha}^2}}{\cosh(\zeta)}
\left[\begin{multlined}
		\sum\limits_{y = 0}^{+\infty}
			\sum\limits_{z = 0}^{y}
			\sum\limits_{q = 0}^{z}
				\left[1 - {\left(1 - \frac{\eta}{4}\right)}^{z-q}\right]
				\left[1 - {\left(1 - \frac{\eta}{4}\right)}^{q}\right] \\
		\abs{\sum\limits_{\substack{0 \leq k_4 \leq z \\ k_4 \equiv y [2]}}
			{t}^{y-z} r^{z-k_4} \sqrt{\binom{y-k_4}{z-k_4}}
			\frac{{(r \alpha)}^{k_4}}{\sqrt{{k_4}!}}
			\sqrt{\binom{y-k_4}{\frac{y-k_4}{2}}}
			{\left[\frac{- \tanh{\zeta}}{2}\right]}^{\frac{y-k_4}{2}}
			\frac{C_{k_4,z-k_4,q}}{\sqrt{2}^z}}^2
  \end{multlined}\right].
\end{equation}

In the previous expression
\begin{equation}
C_{k,l,x}
= \begin{cases}
	\sqrt{\binom{k+l-x}{k} \binom{l}{x}} \hypergeom{-k,-x,l-x+1,-1}                  & \text{ if } 0 \leq \ x \leq l, \\
	{(-1)}^{x-l} \sqrt{\binom{x}{l} \binom{k}{x-l}} \hypergeom{-l,-(k+l-x),x-l+1,-1} & \text{ if } l \leq \ x \leq k+l,
\end{cases}
\end{equation}
with $\hypergeom{p,q,t,w} $ the hypergeometric function.
\end{widetext}

\bibliography{schema_hylight}
% DÉBUT FICHIER .bbl INSÉRÉ
% FIN FICHIER INSÉRÉ

\end{document}